\documentclass{emulateapj}
\usepackage{color} 
\usepackage[colorlinks,urlcolor=blue,citecolor=blue,linkcolor=blue]{hyper ref}

\usepackage{epsfig}
\usepackage{subfigure}
\usepackage{graphicx}
\usepackage{amsmath}
\usepackage{natbib}
\newcommand{\csw}{c_{\mathrm{s,w}}}
\newcommand{\pa}{\partial}
\newcommand{\mb}{\boldsymbol}

\providecommand{\icarus}{Icarus}
\providecommand{\au}{\textsc{au}}

\shorttitle{Magnetized Disk Wind of PPDs}
\shortauthors{Bai, Ye, Goodman \& Yuan}

\begin{document}


\title{Magneto-thermal Disk Wind from Protoplanetary Disks}


\author{Xue-Ning Bai\altaffilmark{1}, Jiani Ye\altaffilmark{2}, Jeremy Goodman\altaffilmark{3}, Feng Yuan\altaffilmark{2}}

\altaffiltext{1}{Institute for Theory and Computation,
Harvard-Smithsonian Center for Astrophysics, 60 Garden St., MS-51, Cambridge, MA 02138}
\altaffiltext{2}{Key Laboratory for Research in Galaxies and Cosmology, Shanghai Astronomical
Observatory,  Chinese  Academy  of Sciences,  80  Nandan Road, Shanghai 200030, China}
\altaffiltext{3}{Department of Astrophysical Sciences, Peyton Hall, Princeton
University, Princeton, NJ 08544}
\email{xbai@cfa.harvard.edu}




\begin{abstract}
Global evolution and dispersal of protoplanetary disks (PPDs) is governed by
disk angular momentum transport and mass-loss processes. Recent
numerical studies suggest that angular momentum transport in the inner
region of PPDs is largely driven by magnetized disk wind, yet the wind mass-loss
rate remains unconstrained. On the other hand, disk mass loss has
conventionally been attributed to photoevaporation, where external heating
on the disk surface drives a thermal wind.
We unify the two scenarios by developing a 1D model of magnetized disk winds
with a simple treatment of thermodynamics as a proxy for external heating.
The wind properties largely depend on 1) the magnetic field strength
at the wind base, characterized by the poloidal Alfv\'en speed $v_{Ap}$,  2) the sound
speed $c_s$ near the wind base, and 3) how rapidly poloidal field lines diverge
(achieve $R^{-2}$ scaling).
When $v_{Ap}\gg c_s$, corotation is enforced near the wind base, resulting in
centrifugal acceleration. Otherwise, the wind is accelerated mainly by
the pressure of the toroidal magnetic field.
In both cases, the dominant role played by magnetic forces likely yields wind
outflow rates that well exceed purely hydrodynamical mechanisms.
For typical PPD accretion-rate and wind-launching conditions, we expect
$v_{Ap}$ to be comparable to $c_s$ at the wind base. The resulting wind is
heavily loaded, with total wind mass loss rate likely reaching a considerable
fraction of wind-driven accretion rate. Implications for modeling global disk
evolution and planet formation are also discussed.
\end{abstract}


\keywords{accretion, accretion disks --- magnetohydrodynamics ---
methods: numerical --- planetary systems: protoplanetary disks}

\section{Introduction}\label{sec:intro}

The theory of magnetized disk winds has been developed for decades, following the seminal
work of \citet[hereafter BP]{BlandfordPayne82}. Assuming a razor-thin (cold) disk and ideal MHD,
poloidal magnetic field lines anchored to the disk behave as rigid wires and will
centrifugally accelerate gas if they are inclined by more than $30^\circ$ to the rotation axis. The
wind exerts a torque on the disk surface that efficiently extracts angular momentum
and drives accretion.
BP's analytical theory was subsequently generalized and refined to better account
for disk structure and boundary conditions, and has been applied to jets and
outflows from general accretion disks and young stellar objects (e.g., 
\citealp{PudritzNorman83,PudritzNorman86,Konigl89,Lovelace_etal91,PelletierPudritz92,
ContopoulosLovelace94,FerreiraPelletier95,Li95,Li96,Ferreira97,Ostriker97,Vlahakis_etal00}).
Global magnetohydrodynamic (MHD) simulations, less restricted by simplifying assumptions
such as self-similarity and time-independence, have
been widely adopted for studying global wind structure and evolution, as well as the
wind launching process and disk-wind connection (e.g., 
\citealp{Ouyed_etal97,OuyedPudritz97a,OuyedPudritz97b,OuyedPudritz99,Ustyugova_etal99,
Krasnopolsky_etal99,Krasnopolsky_etal03,Kato_etal02,FendtCemeIjic02,CasseKeppens02,
CasseKeppens04,Anderson_etal05,Pudritz_etal06,Zanni_etal07,Tzeferacos_etal09,PorthFendt10,
RamseyClarke11,Tzeferacos_etal13}).
It is now well known that the wind properties
depend mainly on the strength and distribution of the magnetic flux threading the disk, and on
the mass loading, with the latter mainly controlled by disk physics and
thermodynamics.

In this paper, we focus on MHD disk winds from very weakly ionized
protoplanetary disks (PPDs). Due to low temperatures,
PPDs are generally extremely weakly
ionized, so that gas and magnetic fields are poorly coupled. 
Except near the star, the coupling depends upon
external ionization sources such as cosmic-rays,
X-rays, and UV photons (e.g., \citealp{Hayashi81,IG99,PerezBeckerChiang11b})
rather than ionization in local thermodynamic equilibrium.
These sources have finite penetration depths and produce
ionization levels that decrease from the disk surface toward the midplane.

It has recently been realized that in the inner region of PPDs ($\lesssim10$ AU),
the magnetorotational instability (MRI, \citealp{BH91}), heretofore supposed to be
the main driver of disk accretion, is almost completely suppressed
\citep{BaiStone13b}. This is mainly because of non-ideal MHD effects.
Ohmic resistivity, the Hall effect, and ambipolar diffusion substantially reduce the
coupling between gas and magnetic fields in weakly ionized gas.
The coupling is in any case too poor for the disk to generate its own field via a self-contained
MRI dynamo \citep{Fleming_etal00,OishiMacLow11,Lesur_etal14,Riols_etal15};
therefore, models for MRI-driven turbulence in PPDs posit net poloidal flux threading the
disk.
Even so, earlier works considering only Ohmic resistivity expected the MRI to operate near the
disk surface ($1-2$ scale heights above midplane, e.g., \citealp{Gammie96},
see also \citealp{Armitage11} for a review).  But ambipolar diffusion
completely changes the picture. Instead of turbulent MRI, the poloidal flux
naturally launches a magnetized disk wind
(\citealp{BaiStone13b,Bai13,Gressel_etal15}, and see
\citealp{Turner_etal14} for a review).
While inclusion of the Hall effect
introduces further complications \citep{Lesur_etal14,Bai14,Bai15,Simon_etal15b},
it appears unavoidable that in the inner regions of PPDs, accretion is
largely wind-driven.

The PPD simulations of \citet{BaiStone13b} and followup works cited above have
demonstrated the robustness of MRI suppression and wind launching.
However, the local (shearing-box) approximations used in these simulations do not permit
the mass-loss and torque of the wind to be reliably determined.
In particular, the location of the Alfv\'en point and mass loss rate depends on the vertical
box size.
Recent simulations
by \citet{Gressel_etal15} were global in radius, but have the
same vertical extent as earlier local simulations, and hence should be
considered local in this sense.

While the aforementioned global studies of MHD disk winds
considered PPDs to be among the primary applications, they
generally failed to take into account realistic disk microphysics.
Furthermore, the range of parameters considered in these studies is not appropriate
for most PPDs, as we elaborate below.

Early analytical wind studies and simulations largely
ignored disk microphysics. While more recent global disk wind simulations take into
account the vertical structure of the disk, they all rely on a prescribed effective magnetic
diffusivity with somewhat arbitrary strength, anisotropy and spatial
dependence. In reality, the diffusivity is due either to MRI turbulence (in well-ionized
disks) and/or to non-ideal MHD effects (in PPDs), effects
that are not captured by these prescriptions. For PPDs, local
simulations that take into account the dominant non-ideal MHD effects indicate
that the wind is launched from several scale heights above midplane, where far-UV
radiation is able to penetrate \citep{BaiStone13b,Bai13,Bai14,Gressel_etal15}.
This location is much higher, and consequently the ratio of the local gas density to that
at the midplane is much lower,  than all previous models assumed.\footnote{Semi-analytical
works of \citet{Konigl_etal10,Salmeron_etal11,Teitler11} also took into account
all non-ideal MHD effects to study wind launching from PPDs, but they did not
consider realistic ionization profile and arrived at different conclusions.}

The choice of wind parameters has usually also been unrealistic, at least for PPDs.
Most previous studies considered very strong disk
magnetization, with midplane strengths close to equipartition:
i.e. $\beta_0\equiv(8\pi p_{\rm gas}/B_z^2)_{\rm midplane}\sim O(1)$, where $p_{\rm gas}$
is the gas pressure and $B_z$ the vertical field.
For standard PPD models, angular momentum transport by such strong field
would be so rapid that the disk would be drained in $\sim10^3$
years instead of typical PPD lifetime of a few Myrs (see Appendix
\ref{app:bstr} for more discussion)\footnote{To achieve realistic wind-driven
accretion rate, wind models constructed by \citet{CombetFerreira08} have
surface density orders of magnitude lower than standard disk models.}.
While near-equipartition field has been suggested to be necessary for wind
launching (e.g., \citealp{WardleKoenigl93,FerreiraPelletier95,Ogilvie12},
partly to suppress the MRI), more recent simulations have showed that with reasonable
prescriptions for magnetic diffusivity, steady disk winds can be launched by much weaker
fields ($\beta_0\sim500$), as long as the disk corona is strongly magnetized
\citep{Murphy_etal10,Sheikhnezami_etal12,StepanovsFendt14}. The
expected level of magnetization in PPDs in standard disk models is still
lower, by several orders of magnitude (see Appendix \ref{app:bstr}).

In addition, most previous MHD wind studies, especially simulations,
adopteded the conventional ``cold MHD wind" scenario, which ignores the thermodynamics
of the wind.   (Notable exceptions are
\citealp{CasseFerreira00b} and \citealp{Tzeferacos_etal13}).
PPDs are quite different from most other
types of accretion disks in that their thermodynamics is dominated by external
heating from the central protostar. In particular, the high-energy stellar X-ray and
UV radiation, which provides the main source of disk ionization, also serves as
the dominant heating source at disk surface and the entire wind zone.
Without considering magnetic fields, such heating is sufficiently strong to drive
``photoevaporation" of PPDs, which has been considered as the dominant
mechanism for PPD mass loss and dispersal  (see recent review by
\citealp{Alexander_etal14}, and also Section \ref{ssec:magpe} for further
discussions). So far, photoevaporation models (pure thermal wind) of PPD mass
loss have been developed in parallel with cold MHD disk wind models without
overlap. In reality, the nature of PPD wind is likely a marriage of both magnetic
and thermal effects: PPD wind is both hot and magnetically dominated.

It is not yet feasible for us to perform global wind simulations that
combine magnetohydrodynamics, tensorial conductivities, chemical and ionization reactions,
and accurate  thermodynamics. Even if we
were able to do so, it might be difficult to interpret the results because they could
be affected by many factors.
Our approach here is to develop a simple global wind model that captures the
essential physics of wind launching, acceleration, and propagation.
The model is an extension of the classic
\citet{WeberDavis67} wind described by relatively few parameters.
This allows us to explore the roles of magnetic and thermal effects easily, if not 
fully consistently, and to reconnoitre those regions of parameter space that
deserve further and more accurate scrutiny.
We also expect our model to serve as a useful framework for interpreting more
realistic global PPD wind simulations in the future.

Our model is similar in spirit to that of \cite[hereafter KS97]{KudohShibata97} in that
we prescribe the poloidal magnetic geometry and solve for the flow, toroidal field, and critical points
along poloidal lines under ideal MHD.  Whereas KS97 were interested
in applications to fast winds launched from $\sim 15\mathrm{R}_{\bigodot}$, however, we are
interested in slower winds launched from $\gtrsim 1\,\au$ where the midplane is
essentially uncoupled to the field, so that the wind base lies several scale
heights above the midplane where X-rays and UV penetrate (Appendix \ref{app:bstr}). Also
we explore a greater range of poloidal magnetic geometries, described by two parameters
rather than KS97's one.  Some of our main conclusions are in agreement with theirs, such
as the importance of toroidal magnetic pressure in lauching the wind.

This paper is organized as follows. In \S\ref{sec:model}, we describe the motivation and
physical picture underlying our wind model. We also describe the wind equations and
parameters, some diagnostic quantities, and our numerical methods. In \S\ref{sec:sol},
after presenting some illustriative wind solutions, we identify the more consequential
wind parameters and systematically explore their influence on wind properties within the
range that we believe relevant to PPDs.  In \S\ref{sec:param}, we show that the important
wind properties, such as mass loading and specific angular momentum, are insensitive to
the remaining input parameters, justifying in part our simplified wind model.  General
implications and broader context are provided in \S\ref{sec:discussion}, and a
summary of our main results and conclusions in \S\ref{sec:sum}.

\section[]{A Simplified Model of Disk Wind from PPDs}\label{sec:model}

We consider the inner region of PPDs ($\lesssim10\,\au$), and assume
axisymmetry and steady state. This is reasonable because
we expect the disk to be largely laminar under realistic conditions
\citep{Bai13,Bai14}. As long as the gas and magnetic field are well coupled
in the wind zone (i.e., in the ideal MHD regime), then gas must flow along
magnetic field lines, and the MHD equations reduce to a set of one-dimensional conservation
laws (e.g., \citealp{Mestel61,Spruit96}, as we will
summarize shortly). This is the basis of MHD wind theory.

In the inner parts of PPDs, the bulk of the gas is cold and extremely weakly ionized.
Non-ideal MHD effects (particularly Ohmic resistivity
and ambipolar diffusion) are strong within several scale heights of the midplane
(e.g., \citealp{Wardle07,Bai11a,Mohanty_etal13}). As a result, gas is
very poorly coupled to magnetic fields, and does not follow field lines, so that an
MHD wind can not be launched efficiently. On the other hand,
FUV can almost fully ionize trace species such as carbon and sulfur in the disk upper
atmosphere, down to a column density $\sim0.01-0.1$g cm$^{-2}$,
providing ionization fractions $\gtrsim10^{-4}$ \citep{PerezBeckerChiang11b}. At this
ionization level, non-ideal MHD effects become relatively unimportant in wind dynamics
(but perhaps not thermodynamics: e.g., \citealp{Safier93,Garcia_etal01}).
Indeed, with a simplified prescription for
FUV ionization, local simulations by \citet{BaiStone13b} showed that the base of the
wind\footnote{Technically, this corresponds to the location 
where the gas velocity transitions from sub-Keplerian (below) to
super-Keplerian (above), marking the onset of centrifugal acceleration
\citep{WardleKoenigl93}. While the wind launching process starts from
somewhat deeper layers where coupling between gas and magnetic fields
is marginal, the wind base so defined marks a clear transition in the flow
properties (see Section 4 of \citealp{BaiStone13b}).}
generally coincides with the FUV ionization front. The same conclusion is
also arrived from (radially) global simulations of \citet{Gressel_etal15}.

The FUV radiation not only significantly enhances the ionization fraction, but also
significantly heats the gas. The thermal structure of the disk surface is highly complex,
with differing gas and dust temperatures, strong molecular and atomic line emission,
and other non-LTE effects (see \citealp{Bergin_etal07} for a review). It has been
extensively studied in the framework of hydrostatic disk models (e.g.,
\citealp{Glassgold_etal04,NomuraMillar05,AikawaNomura06,Woitke_etal09,Walsh_etal10,
BethellBergin11,Fogel_etal11,Walsh_etal12,Akimkin_etal13}).
In general, the gas temperature increases dramatically from the cold disk through the CO
photodissociation front and reaches a few thousand degrees in the FUV layer.

\subsection[]{Model Description}\label{ssec:dscp}

\begin{figure}
    \centering
    \includegraphics[width=90mm]{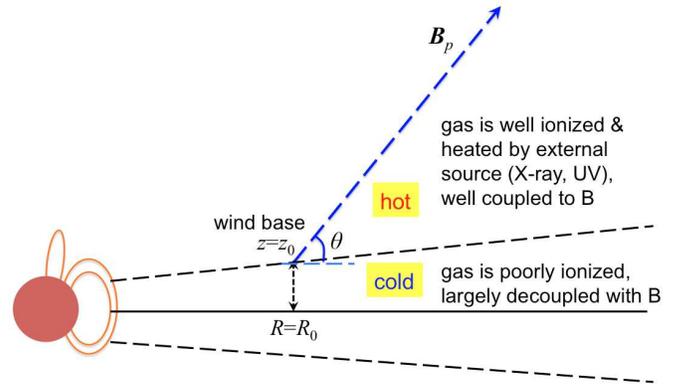}
  \caption{Schematic illustration of our simplified wind model. See Section
  \ref{ssec:dscp} for details.}\label{fig:illus}
\end{figure}

Motivated by these discussions, we now describe a simplified
magnetized model for PPDs as illustrated in Figure \ref{fig:illus}. We use
$R$ to denote cylindrical radius, and consider disk wind launched from a
fiducial radius $R_0$. The disk is divided into a cold interior, where the
gas is poorly coupled to magnetic fields and obeys hydrostatic equilibrium,
and a hot exterior, where the gas is sufficiently well coupled
and an MHD wind is launched. The transition occurs at vertical height
$z=z_0$ from disk midplane, the location of the wind base.
We prescribe the geometry of a typical poloidal field line anchored to the
disk at the wind base ($R_0, z_0$) and solve the conservation laws along
this wind field line. For simplicity, poloidal field lines are taken to be
straight, with a constant inclination angle $\theta$ relative to the equatorial
plane at $z\ge z_0$. In parametric form, the field line is described by
\begin{equation}
R(s)=R_0+s\cos\theta\ ,\qquad z(s)=z_0+s\sin\theta\ ,\label{eq:wline}
\end{equation}
where $s$ is the arc length along the poloidal field line.

In addition to poloidal field geometry, we also need to prescribe the functional
form of $B_p(R)$. If the wind is launched from a radially extended region in
an approximately self-similar manner, then the poloidal field would
be approximately parallel near the wind launching region, and hence
$B_p(R)\propto R^{-1}$. Towards large $R$, the solenoidal constraint
demands $B_p(R)\propto R^{-2}$. (This is true of KS97's field, for example,
although it collimates parabolically toward the axis, unlike ours.)
We therefore assume the following functional form for $B_p(R)$:
\begin{equation}
B_p(R)=B_{p0}\frac{1+q}{(R/R_0)+q(R/R_0)^2}\,.\label{eq:bpR}
\end{equation}
The parameter $q$ controls the radius at which the poloidal field lines
transition from being parallel to diverging.

Along the field line, let $\rho$, ${\mb v}_p$, and $v_\phi=\Omega R$
be gas density, poloidal  velocity, and toroidal velocities, with $\Omega$ the
gas angular velocity. Poloidal and toroidal magnetic fields are denoted
${\mb B}_p$ (prescribed) and $B_\phi$ (to be solved for).  As noted
earlier, the ideal MHD equations for a steady axisymmetric wind
yield four constants of the flow.  The first three of these are
\begin{equation}
k\equiv\frac{4\pi\rho v_p}{B_p}\,,\label{eq:mas}
\end{equation}
which follows from the continuity equation,
\begin{equation}
\omega\equiv\Omega-\frac{kB_\phi}{4\pi\rho R}\,,\label{eq:omg}
\end{equation}
which follows from the induction equation and can be interpreted
as the angular velocity of the line, and the specific angular momentum
\begin{equation}
l=\Omega R^2-\frac{RB_\phi}{k}\,.\label{eq:ang}
\end{equation}

Thermodynamics is reflected in the energy equation via the enthalpy
$h\equiv\int dP/\rho$. We assume that the gas in the wind zone can be
approximately described by a barotropic equation of state $P=P(\rho)$
so that we simply have $h=h(\rho)$. By default, we choose the wind to
be isothermal at sound speed $\csw$, so that
$P=\rho\csw^2$ and
\begin{equation}\label{eq:hiso}
h(\rho)=\csw^2\log{(\rho/\rho_0)}\ ,
\end{equation}
where $\rho_0$ is the density
at the wind base. 
Energy conservation is then expressed via the Bernoulli constant
\begin{equation}
E=\frac{v^2}{2}+h+\Phi-\omega Rv_\phi
=\frac{v_p^2+(v_\phi-\omega R)^2}{2}+h+\Phi_{\rm eff}\ ,\label{eq:eng}
\end{equation}
where $v=\sqrt{v_p^2+v_\phi^2}$ is the total flow speed,
$\Phi=-GM_*/\sqrt{R^2+z^2}$ is the gravitational
potential, and $\Phi_{\rm eff}=\Phi-\tfrac{1}{2}\omega^2R^2$.
Unlike ``viscous'' angular momentum transport (e.g., MRI
turbulence), ideal MHD winds do not directly heat the disk.\footnote{A low level
of heating can be produced as magnetic flux drifts through the gas in the disk
interior---and perhaps even within the wind \citep{Safier93}.} 

Although constant along each poloidal line,
the four flow quantities $\{k,\omega,l,E\}$ vary between lines.
Once we have prescribed the shape of a poloidal field line
and its strength $B_p$ along the line, as well as
the enthalpy $h(\rho)$, the wind solution is fully specified by the flow constants.
The primitive flow variables ($\{\rho, v_p, v_\phi, B_\phi\}$) are determined
by $\{k,\omega,l,E\}$ via the defining equations \eqref{eq:mas}-\eqref{eq:eng}.
However, not all of the flow constants can be chosen independently if the
wind is to reach infinity with a terminal speed exceeding all of its characteristic velocities.
The Bernoulli constant $E$ is essentially determined by physical conditions at
the wind base.  Also, one expects
$\omega\approx\Omega_K(R_0)$, where $\Omega_K(R_0)$ is the Keplerian
angular frequency at the wind base, and so we provisionally take $\omega=\Omega_K$.
The remaining constants $k$ and $l$ are found by imposing
regularity conditions at two critical points.

For a given disk model, therefore, the wind solution is determined by 
the following parameters:
\begin{itemize}
\item Physical parameters: $B_{p0}$ and $q$ [eq.~\eqref{eq:bpR}] and $\csw$ [eq.~\eqref{eq:hiso}].
\item Geometric parameters: base height $z_0$ and field inclination angle $\theta$.
\end{itemize}
We provide the expected range of these parameters and normalize
them to dimensionless units in \S\ref{ssec:units}. The more general
situation that $\omega\neq\Omega_K$ is discussed in \S\ref{ssec:omega}.

Needless to say, our wind model involves substantial simplifications, especially with
regard to the poloidal field and thermodynamics. In reality, the poloidal field geometry
and strength are determined by force balance perpendicular to flux surfaces, as described
by the Grad-Shafranov equation (GSE), a nonlinear partial differential equation in $R$ and
$z$. Besides the numerical difficulties involved in finding it, the solution to this equation
depends on the global distribution of magnetic flux, which is largely unconstrained {\it a
  priori}. As a result, even when self-similarity is imposed, a diverse set of wind solutions is
possible with vastly different field geometries. For instance, \citet{Ostriker97} showed
that cylindrical collimation of a self-similar wind will not be achieved if the radial
gradient of gas density and magnetic flux is steep, as is expected for PPDs.

Therefore, rather than solve the GSE, we have chosen to
parametrize the poloidal field in terms of $B_{p0}$ (its strength at the wind
base), $\theta$ (inclination to the disk), and $q$ (divergence radius $q^{-1}R_0$).
We expect the effects of more realistic field geometries to be roughly
captured by systematically varying $\theta$ and $q$.
Similarly, the role of external heating is encapsulated by an isothermal sound speed
$\csw$, and by the wind-base height $z_0$, which reflects the
penetration depth of external radiation.
We also consider a polytropic equation of state in \S\ref{ssec:eos} to
further test our simplified thermodynamics.
In brief, despite the simplicity of our model, its parameters accomodate
the essential physics of magnetized disk winds, and its simplicity
facilitates the exploration and interpretion of their properties.

\subsection[]{Properties of Wind Equations and Critical Points}\label{ssec:crit}

It is convenient to define
\begin{equation}
x\equiv\frac{4\pi\rho}{k^2}=\frac{v_{Ap}^2}{v_p^2}\ ,
\end{equation}
which rescales gas density $\rho$ and marks the ratio of the poloidal Alfv\'en
velocity $v_{Ap}=B_p/\sqrt{4\pi\rho}$ to flow velocity.  Eliminating
$B_\phi$ from Equations (\ref{eq:omg}) and (\ref{eq:ang}) yields
\begin{equation}
\omega-\Omega=\frac{l-\omega R^2}{(x-1)R^2}=\omega\frac{(R_A/R)^2-1}{x-1}\ .
\label{eq:alfven}
\end{equation}
This relation defines the Alfv\'en point/radius ($R=R_A$): when the wind
poloidal velocity reaches the Alfv\'en speed ($x=1$), the numerator must also vanish,
whence
\begin{equation}
l=\omega R_A^2\ .\label{eq:RA}
\end{equation}
The ratio $R_A/R_0$ is often called the magnetic lever arm, and
$(R_A/R_0)^2$ is commonly denoted by $\lambda$ following BP.
Although singular, the Alfv\'en point does not yield any
additional constraint on the wind solution, and hence is not a critical point of the
problem\footnote{The Alfv\'en point does become a critical
point in solving cross-field balance via the GSE
(e.g. BP, \citealp{Spruit96}).}.

Eliminating $v_p$ and $v_\phi=\Omega r$ from 
the energy equation (\ref{eq:eng}) via
eqs.~(\ref{eq:mas}) \& (\ref{eq:alfven}) yields
\begin{equation}
\begin{split}
E&=\frac{1}{2}\frac{B_p^2}{k^2x^2}
+\frac{\omega^2R^2}{2}\frac{(R_A^2/R^2-1)^2}{(x-1)^2}+h+\Phi_{\rm eff}\\
&\equiv H(x,s)\ .\label{eq:E2}
\end{split}
\end{equation}
Once the values of $k$, $R_A$, \& $\omega$ are known, one can
obtain  $E=H(x,s)|_{\rho=\rho_0, s=0}$ and then solve the algebraic equation
$H(x,s)=E$ to obtain $x$ (or $\rho$) as a
function of position along the line ($s$, or $R$), thereby completing the wind solution.

For nonzero sound speed, however,
additional constraints among $k$, $R_A$, $\omega$, and $E$
must be satisfied for a smooth solution.
This can be seen from
\begin{equation}
\begin{split}
x\frac{\pa H}{\pa x}&=-v_p^2-\frac{x}{x-1}(\Omega-\omega)^2R^2+\csw^2\\
&=\frac{v_p^4-(\csw^2+v_A^2)v_p^2+\csw^2v_{Ap}^2}{v_{Ap}^2-v_p^2}\ ,
\end{split}
\end{equation}
where $v_A=\sqrt{(B_p^2+B_\phi^2)/4\pi\rho}$ is the total Alfv\'en velocity.
Clearly, the numerator becomes zero when the poloidal wind velocity $v_p$
equals the poloidal component of the slow or fast magnetosonic speed
($v_{sp}$ or $v_{fp}$),
\begin{equation}
v_{sp}^2, v_{fp}^2=\frac{(\csw^2+v_A^2)\mp\sqrt{(\csw^2+v_A^2)^2-4\csw^4}}{2}\ .
\end{equation}
Correspondingly, $\pa H/\pa x=0$ defines the slow and fast magnetosonic
points.
In general, the algebraic equation $H(x,s)=E$ either has no solution for $x$ at each $s$ or
two disconnected branches of solutions. A smooth wind solution is
possible only when different branches join at the slow
and fast magnetosonic points. At each of these two critical points, the following relations must
be satisfied:
\begin{equation}\label{eq:crit}
H(x,s)=E\ ,\quad\frac{\pa H(x,s)}{\pa x}=0\ ,\quad\frac{\pa H(x,s)}{\pa s}=0\ .
\end{equation}
Since these are three equations for two variables, a
constraint results at each critical point, thereby determining $k$ and $R_A$
(equivalently, $k$ \& $l$) in terms of $\omega$ \& $E$.

In sum, to obtain the wind solution, the three algebraic equations
(\ref{eq:crit}) must be solved at both slow and fast magnetosonic points,
which yield the location (radius) and density of the two critical points
($R_s$, $x_s$, $R_f$, $x_f$), together with mass flux $k$ and Alfv\'en radius
$R_A$. Energy flux is implicitly determined from $E=H(x,s)|_{\rho=\rho_0, s=0}$.

These equations apply to any axisymmetric, steady-state,
ideal-MHD wind. For instance, \citet{WeberDavis67}'s classic model
is the special case $\{\theta=0,\,q=\infty,\,z_0=0\}$.
Our wind solutions share many similarities with theirs.  But with their
inclined field geometry and launching from the disk surface, our winds
experience different $\Phi_{\rm eff}$, leading to different energetics.

\subsection[]{Dimensionless Units and Model Parameters}\label{ssec:units}

In our calculations, we normalize lengths, velocities, and gas densities by their values at
the footpoint of the field-line on the disk (wind base), so that $R_0=v_K=\rho_0=1$. Correspondingly,
time is in units of $1/\Omega_K$.

Poloidal field strength is conveniently parameterized by
$v_{A0}\equiv B_{p0}/\sqrt{4\pi\rho_0}$, the poloidal Alfv\'en velocity at the
footpoint. Appendix \ref{app:bstr} discusses the expected parameter
values for winds launched from $R_0\sim 1\,\au$.  Based on that
discussion, we choose the following fiducial parameters:
\begin{equation}
\begin{split}
&v_{A0}=\csw=0.1v_K\ ,\\
&z_0=0.15R_0\ ,\quad \theta=45^\circ\ ,
\end{split}
\end{equation}
with $\omega=\Omega_K$. We take $q=0.25$ as our fiducial choice for the
functional form of $B_p$ in Equation (\ref{eq:bpR}), which means that
transition from $R^{-1}$ (parallel) to $R^{-2}$ (diverging) occurs
at around $R\sim2R_0$. 

We find that $q$, $v_{A0}$ and $\csw$ are the more important model parameters
and focus on the dependence of wind solutions on these parameters in
\S\ref{sec:sol}.  We vary $v_{A0}/v_K$ in the range
between $0.01$ and $10$, and $\csw/v_K$ between $0.03$ and $0.3$.
When exploring the role of poloidal field strength,
only $v_{A0}$ needs to be varied. However, when exploring the role of
wind temperature at fixed $B_{p0}$, in addition to varying $\csw$,
one must also vary $v_{A0}$ proportionally. This is because at fixed
pressure (at wind base), the density scales inversely with temperature, thus
affecting the Alfv\'en speed. For the field line divergence parameter $q$, we
also consider $q=0.1$ (where transition occurs at $\sim3R_0$)
and $q=\infty$ (purely diverging).

The dependence on the remaining parameters ($\theta$, $z_0$, $\omega/\Omega_K$, equation
of state) is explored in \S\ref{sec:param} and found to be relatively weak.

\subsection[]{Diagnostics of Wind Kinematics}\label{ssec:diag}

The most important wind diagnostic is the Alfv\'en radius $R_A$ (i.e. $R_A/R_0$): this
characterizes the efficiency of the wind for extracting disk angular momentum.  Since
$\omega\approx\Omega_K$, the excess angular momentum per unit mass in the wind is
$\Omega_K(R_A^2-R_0^2)$.  The removal of this excess from the disk leads to a direct
relation between the wind mass loss rate and wind-driven accretion rate.

Let $\dot{M}_{\rm acc}$ be the wind-driven accretion rate at radius $R_0$.  
Angular-momentum conservation leads to
\begin{equation}\label{eq:mdot1}
\dot{M}_{\rm acc}\frac{dj}{dR}=\frac{d\dot{M}_{\rm wind}}{dR}\Omega_K(R_A^2-R_0^2)\ ,
\end{equation}
where $j(R)\equiv\Omega_KR^2$ is the specific angular momentum in the disk,
and $\dot{M}_{\rm wind}(R)$ denotes the cumulative mass loss rate from the origin to
disk radius $R$. Noting that $d\ln j/d\ln R=1/2$ in Keplerian disks, we obtain
\begin{equation}
\xi\equiv\frac{d\dot{M}_{\rm wind}/d\ln R}{\dot{M}_{\rm acc}}\bigg|_{R=R_0}
=\frac{1}{2}\frac{1}{(R_A/R_0)^2-1}\ .\label{eq:mdot2}
\end{equation}
\cite{FerreiraPelletier95} call $\xi$ the ``ejection index."

\begin{figure*}
    \centering
    \includegraphics[width=150mm]{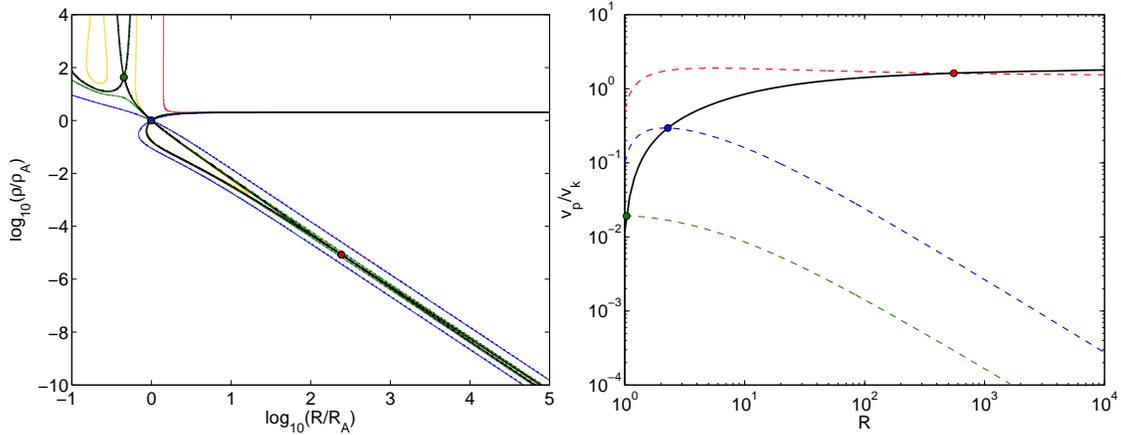}
    \caption{Representative wind solution for our fiducial parameters $q=0.25$,
      $v_{A0}=\csw=0.1v_K$, $\omega=\Omega_K$, $\theta=45^\circ$, $z_0=0.15R_0$.
      \emph{Left:} contours of the energy function $H(\rho, R)$. Black solid lines
      correspond to $H=E$, the Bernoulli constant, for the solution whose branches connect
      smoothly at both critical points.  Yellow and red contours correspond to
      $H=E-0.2v_K^2$, and $H=E-4v_K^2$, respectively; green and blue contours to
      $H=E+0.2v_K^2$, and $H=E+4v_K^2$. \emph{Right:} poloidal velocity profile of the
      wind solution (solid), and the green, blue and red dashed lines mark the poloidal
      components of slow magnetosonic, Alfv\'en and fast magnetosonic velocities. In both
      panels, green, blue and red dots mark the location of the slow, Alfv\'en and fast
      points.  }\label{fig:contour}
\end{figure*}

Another important diagnostic quantity is the dimensionless ``mass loading"
parameter:
\begin{equation}\label{eq:mu}
\mu\equiv\frac{\omega R_0}{B_{p0}}k=\frac{\omega R_0v_{p0}}{v_{A0}^2}\ .
\end{equation}
A wind is considered ``heavily loaded" if $\mu\gtrsim1$,  and
``lightly loaded" if $\mu\ll1$.

Mass accretion is due to the magnetic torque exerted at the disk
surface/wind base (e.g. \citealp{BaiGoodman09})
\begin{equation}\label{eq:diracc}
\dot{M}_{\rm acc}=\frac{2R_0}{\Omega_K}|B_zB_{\phi}|_{z_0}
=\frac{2R_0}{\Omega_K}B^2_{p0}|\frac{B_{\phi}}{B_{p}}|_{z_0}\sin\theta\ .
\end{equation}
The mass loss rate per logarithmic radius 
\begin{equation}
\frac{d\dot{M}_{\rm wind}}{d\ln R}=2\pi R^2(\rho v_p)_{z_0}
=2\pi R^2\rho_0v_{p0}\, ,
\end{equation}
can be rewritten using (\ref{eq:mu}) as
\begin{equation}\label{eq:dirmloss}
\frac{d\dot{M}_{\rm wind}}{d\ln R}=\frac{2\pi R_0}{\omega}\mu\rho_0v_{A0}^2
=\frac{R_0}{2\omega}\mu B_{p0}^2\ .
\end{equation}
We see that accretion rate and mass loss rate have a common scaling factor
$R_0B_{p0}^2$. With $\omega\approx\Omega_K$, their ratio is then
determined by the values of $|B_{\phi}/B_p|_{z0}$ and $\mu$, which are determined by the
wind solution.\footnote{Strictly, eq.~(\ref{eq:diracc}) assumes that the
wind base occurs where $v_\phi=\Omega_KR_0$
\citep{WardleKoenigl93}. In our model, since $v_\phi$ is obtained from the wind
solution, it does not strictly equal $\Omega_KR_0$ at our wind base.
However, in practice the deviation is typically very small, and the ratio
of (\ref{eq:diracc}) to (\ref{eq:dirmloss}) agrees with (\ref{eq:mdot2}) within $10\%$.
In (vertically-local) simulations \citep{BaiStone13b,Gressel_etal15}, the wind
base as defined by WK93
coincides closely with the FUV ionization front (which we adopt as the wind base).}

Other wind diagnostics are also of interest. One is the pitch of the field
at the Alfv\'en point,
\begin{equation}
\frac{-B_\phi}{B_p}\bigg|_{R_A}=\frac{\omega R_A}{v_{Ap}}
\bigg(-\frac{d\ln\rho}{d\ln R}\bigg)\bigg|_{R_A}^{-1}\,,
\end{equation}
which is a useful indicator of wind acceleration mechanism.
Another is the wind terminal velocity
$v_p^{\rm inf}$, which relate the wind momentum and kinetic
energy flux.
A third is the ratio of Poynting flux to kinetic energy flux $\sigma$,
which measures the efficiency of conversion of magnetic to kinetic energy.
From Equations (\ref{eq:mas}) and (\ref{eq:omg}), it is straightforward to
show that the Poynting flux is $-\omega RB_pB_\phi/4\pi$.
At $R\gg R_A$, we expect $\Omega\sim\omega R_A^2/R^2\ll\omega$,
hence from Equations (\ref{eq:omg}), we have
$B_\phi\approx B_p(\omega R/v_p)$. With these relations, we obtain
\begin{equation}
\sigma=\frac{-\omega RB_pB_\phi}{2\pi\rho v_p^3}\approx
\frac{B_\phi^2}{2\pi\rho v_p^2}\approx2\frac{v_A^2}{v_p^2}\ .\label{eq:sigma}
\end{equation}
We note that because our wind solution passes the fast magnetosonic
point, and $v_{fp}>v_A$, the asymptotic value of $\sigma$ at infinity
must be smaller than $2$.

For a cold wind ($c_s\sim0$) in the Weber \& Davis wind model
($\theta=0$, $q=\infty$), \citet{Spruit96} showed that the slow and fast
magnetosonic points are located at the wind footpoint and infinity,
respectively, and obtained analytical wind solutions. In particular, a very
useful relation derives from the fast-point condition:
\begin{equation}
\bigg(\frac{R_A}{R_0}\bigg)^2=\frac{3}{2}(1+\mu^{-2/3})\ .\label{eq:mura}
\end{equation}
This scaling relation has been commonly observed in simulations cold
MHD disk winds \citep{OuyedPudritz97a,Anderson_etal05,Zanni_etal07}.
Another useful relation states that
\begin{equation}
v_p^{\rm inf}/(\omega R_0)=\mu^{-1/3}\,.\label{eq:muvp}
\end{equation}
Hence lightly loaded winds achieve higher terminal
velocities, although limited by the weak $1/3$ power.
Spruit also showed that at the Alfv\'en point,
\begin{equation}\label{eq:mubphi}
\bigg|\frac{B_{\phi}}{B_p}\bigg|_{R_A}\approx
\begin{cases}
(19/8)^{1/2} & (\mu\ll1) \\
1.14\mu & (\mu\gg1)
\end{cases}\,.
\end{equation}
Hence heavily-loaded winds have tightly wrapped toroidal fields
at the Alfv\'en point.
We will test these relations in our more general wind models.

\begin{figure*}
    \centering
    \includegraphics[width=183mm]{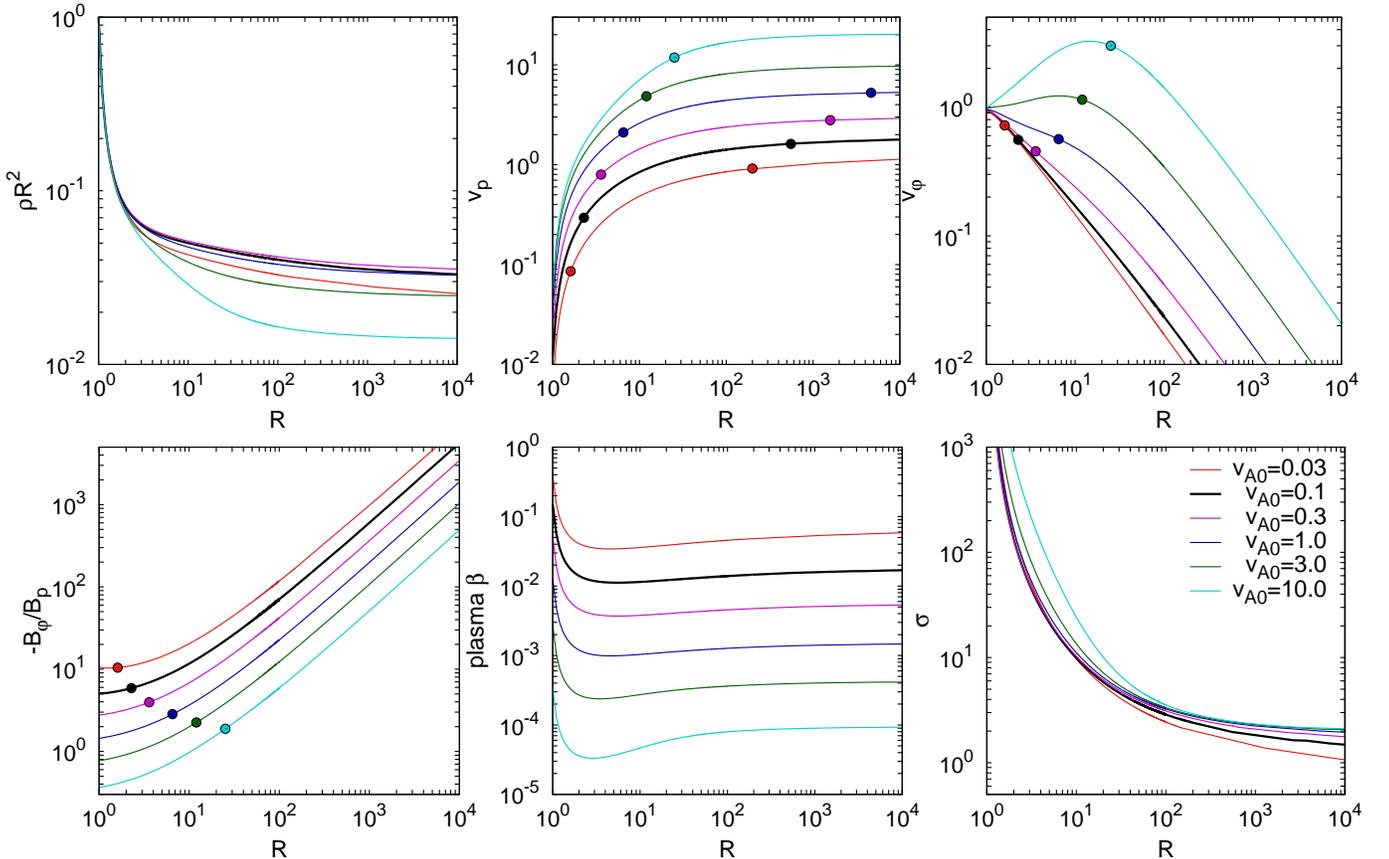}
    \caption{Profiles of various physical quantities along poloidal field lines as
      functions of cylindrical radius $R$.  We vary field strength $v_{A0}$ (see legend)
      while fixing $\csw=0.1v_K$, $q=0.25$.  Bold black line corresponds to our
      fiducial set of parameters.  Alfv\'en points of the solutions are marked by
      filled circles in three of the panels.   Fast magnetosonic points are marked in
      top middle panel (but lie outside the box for two top curves).
      {\it Bottom middle panel:} $\beta$ is the ratio of gas pressure to magnetic
      pressure, computed from the total field.  {\it Bottom right panel:} $\sigma$ is the
      ratio of Poynting flux to kinetic energy flux (eq.~\ref{eq:sigma}).}\label{fig:prof}
\end{figure*}

\subsection[]{Numerical Method}

We solve the six algebraic equations in Equations (\ref{eq:crit}) by a series
of numerical iterations. In brief, we start from initial guesses for $R_A$ and
$k$ and solve for the locations of the slow and fast magnetosonic points
by $\pa H/\pa\ln x=\pa H/\pa\ln R=0$. We analytically calculate the second
derivatives of $H$ with respect to $x$ and $R$ to guide the root-finding
routines. The next stage of iteration involves adjustments of $k$ and $R_A$
so that $H(x_s,R_s)=E$ and $H(x_f,R_f)=E$ are satisfied, via a version of
Newton's method.

To illustrate the mathematical structure of a typical wind solution, we show
in Figure \ref{fig:contour} for our fiducial parameter set
the contours of the energy function $H$ in the plane of
$\rho/\rho_A$ and $R/R_A$.
Clearly, the slow and fast critical points are
saddle points of $H$. A smooth wind solution requires that these two saddle
points lie at intersections of contours where $H=E$, the Bernoulli constant. The solution
for $\rho$ (or $x$) as a function of $R$ (or $s$) smoothly connects the three
singular points (slow, Alfv\'en and fast). Beyond the fast magnetosonic point,
the solution takes the lower branch. The right panel of the Figure further
illustrates that the flow is continuously accelerated, passes the
three MHD wave velocities at the location of the three singular points, and
achieves an asymptotic velocity that is close to the Keplerian velocity at
the wind launching point.

While solving for the slow point and $k$ at fixed $R_A$ is straightforward, success in
finding the fast point sometimes requires a shrewd initial guess. This is related to the
fact that the two branches of the $H=E$ contour are almost parallel to each other near the
fast point (Fig.~\ref{fig:contour}).  We surmount this numerical difficulty by starting
from a standard wind solution, and then gradually altering its parameters towards the desired set,
using each solution as the initial guess for the next.  Once $k$ and $R_A$ are found, it
is straightforward to obtain the solution $x=x(s)$ by solving $H(x,s)=E$ either by
iteration or by tracking the $H=E$ contours from the critical points.

In certain cases, the slow point is located below the wind
base with $R_s<R_0$, a situation that violates our assumption that ideal
MHD applies throughout the wind (including the critical points). When this
occurs, our wind solution is inaccurate, and we mark it separately.

\begin{figure*}
    \centering
    \includegraphics[width=183mm]{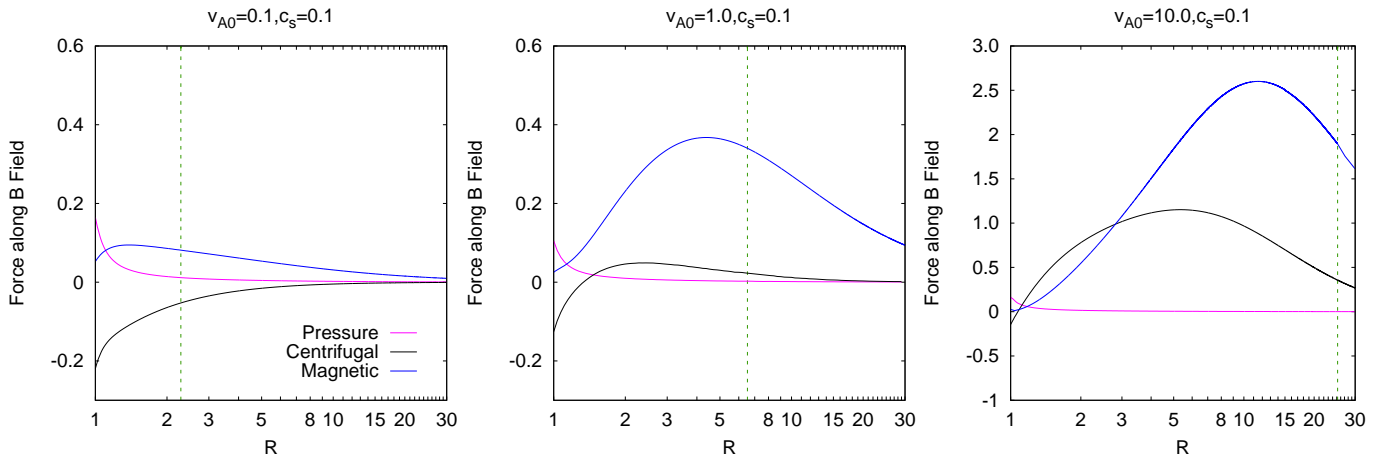}
  \caption{Wind acceleration mechanism for three of the
  solutions shown in Fig.~(\ref{fig:prof}) with $v_{A0}/v_K=0.1$,
  $1.0$ and $10.0$. Poloidal forces are decomposed
  into pressure-gradient, centrifugal, and magnetic according
  to eq.~(\ref{eq:accmech}).  Vertical dashed green lines
  mark the Alfv\'en radii.}\label{fig:force}
\end{figure*}

\section[]{Representative Wind Solutions}\label{sec:sol}

In this section, we present representative wind solutions where we fix the
wind geometry at $\theta=45^\circ$ at $z_0=0.15R_0$, and choose
$\omega=\Omega_K$. We explore the role played by key parameters
including field strength ($v_{A0}$), field divergence parameter $q$, and
wind temperature ($\csw$), varying them around our fiducial values
$v_{A0}=\csw=0.1v_K$ and $q=0.25$.

\subsection[]{Role of Poloidal Field Strength}\label{ssec:Bpstr}

We first fix $\csw=0.1$ and $q=0.25$, and explore the role of poloidal
field strength by varying $v_{A0}/v_K$ from $0.03$ to $10$. In Figure
\ref{fig:prof}, we show the radial profiles of various wind diagnostic quantities.

All solutions accelerate monotonically from the wind-launching region, smoothly pass the
critical points, and approach terminal velocities comparable to or beyond the Keplerian
velocity at the wind base. Typically, near the wind base, the poloidal velocity is very
small, and the gas is pressure supported, with significant contribution from magnetic
pressure (see \S\ref{sssec:mechanism}).
In all solutions, the gas density decreases with $R$ (and hence $z$) as if in
(magneto-)hydrostatic equilibrium in this region. 
At larger radii, the wind achieves its asymptotic velocity $v_p^{\rm inf}$, and hence the
gas density scales in the same way as $B_p$, i.e. $R^{-2}$, while the scalings
$B_\phi\propto R^{-1}$ and $v_\phi\propto R^{-1}$ follow directly from the conservation
laws.

The Alfv\'en radius $R_A$ is a rapidly increasing function of $v_{A0}$.  Using a small
$v_{A0}=0.03v_K$, we obtain $R_A\approx1.61R_0$. 
Such small $R_A$ would result in extremely efficient mass loss, with ejection index
$\xi=(d\dot{M}_w/d\ln R)/\dot{M}_a\approx0.4$ (eq.~\ref{eq:mdot2}). If the wind were
launched over an extended region of the disk---say $0.3$-$15\,\au$ as suggested by local
disk simulations \citep{Bai13}---the integrated wind mass loss rate (assuming $R_A/R_0$
roughly constant) would even exceed the accretion rate: $\dot{M}_w\approx1.5\dot{M}_a$! We
defer the discussion of the possible consequences of such extreme mass loss to \S
\ref{sec:discussion}. On the other hand, increasing $v_{A0}$ to values $\gtrsim v_K$ moves
the Alfv\'en radius beyond $5R_0$.  In this limit, we have $\xi\lesssim0.02$ and mass loss
is almost negligible compared with accretion. Our fiducial solution ($v_{A0}=\csw=0.1v_K$)
has $R_A/R_0\approx2.288$, gives a modest ejection index $\xi\approx0.12$ which, if
approximately constant over an extended radial range (e.g., again $0.3$-$15\,\au$), still
yields a total mass loss rate that is a considerable fraction ($47\%$) of the wind-driven
accretion rate.

For a cold ($\csw\ll v_{A}$) wind, \citet{Ferreira97} derived the upper limit $\xi<0.15$
from energetic considerations. We are not restricted by this constraint because our wind
has finite temperature.  Our solutions violate this relation mainly when
$v_{A0}\lesssim\csw$, so that the wind is not cold in the launching region.  Later in
\S\ref{ssec:temp}, we will see that over a wide range of parameter space, it is the ratio
$v_{A0}/\csw$ that largely sets the value of $R_A$ and hence $\xi$.

The fast magnetosonic points are generally located at very large distances.  Although they
are needed to find the Alfv\'en radii, the general wind
properties are insensitive to the conditions at large distances (see \S\ref{ssec:eos} for
an example). This is because the Bernoulli constant (\ref{eq:eng}) is largely dominated by
the kinetic energy terms there, while the enthalpy and gravitational potential terms
become negligible.

In all cases, the wind is magnetically dominated throughout, as measured
by plasma $\beta$, defined as the ratio of gas pressure to total magnetic
pressure. Asymptotically, toroidal field wind up, giving $B_\phi/B_p\propto R$.
At the wind base, toroidal field still dominates unless the system is very
strongly magnetized with $v_{A0}\gtrsim v_K$, as seen from the bottom
left panel of Figure \ref{fig:prof}. This fact is directly related to the wind
acceleration mechanism, as we discuss below in detail.

\subsubsection[]{Acceleration mechanism}\label{sssec:mechanism}

The wind acceleration process is best understood by directly considering
the poloidal forces:
\begin{equation}\label{eq:accmech}
\frac{dv_p}{dt}=\bigg[-\frac{dh}{dR}
+\bigg(\frac{v_\phi^2}{R}-\frac{d\Phi}{dR}\bigg)
-\frac{B_\phi}{4\pi\rho R}\frac{d(RB_\phi)}{dR}\bigg]\frac{dR}{ds}\,.
\end{equation}
The first term in brackets is the thermal pressure gradient, and the
last combines the tension and pressure gradient of the toroidal field.

We define the second term as the net centrifugal force (per unit mass), meaning the excess
of centrifugal over gravitational acceleration.  In the classical picture of
magneto-centrifugal winds, a strong poloidal field dominates in the wind launching
region. The stiffness of the poloidal field forces the flow to corotate with the wind
footpoint (base) up to the Alfv\'en point, and accelerates the gas like
``beads-on-a-wire." The net centrifugal force thus dominates the acceleration in this
scenario.

In Figure \ref{fig:force}, we show the force decomposition discussed
above for three representative wind solutions shown in Figure
\ref{fig:prof} with different $v_{A0}$. We see that when $v_{A0}$ is
small ($\sim0.1v_K$), the net centrifugal force is always negative,
and hence contributes negatively to the acceleration, which
is almost entirely driven by the magnetic pressure gradient. 
The dominant role of magnetic pressure gradient in the case of weak poloidal field
was also pointed out in KS97, and discussed in \citet{LyndenBell96,LyndenBell03}.
For larger $v_{A0}\sim v_K$, the net centrifugal force is positive over a finite range of
radii, yet acceleration is still dominated by magnetic forces.
Only when $v_{A0}\gg v_K$ does centrifugal acceleration dominate, although
the dominance never extends to the Alfv\'en point within our range of parameters.

The effectiveness of centrifugal acceleration is directly related to
how well co-rotation can be enforced. This can be seen from the
top right panel of Figure \ref{fig:prof}: co-rotation is
enforced only for $v_{A0}=10v_K$.
For smaller $v_{A0}$, the gas rotates substantially more slowly, and when
$v_{A0}\sim \csw$, $v_{\phi}$ begins almost immediately to drop as
$1/R$, suggesting little exchange of angular momentum with the field.

The lack of co-rotation is in line with the dominance of toroidal field discussed
earlier. Enforced co-rotation only when the poloidal field is stiff enough to resist the
tendency of the inertia of the gas to bend it into toroidal field.  The fact that our wind
is not ``cold" makes magnetic field more prone to bending by increasing the mass
loading.

In brief, our results suggest magnetic pressure gradient to plays a
dominant role on wind launching in protoplanetary disks.
In their original local study of PPD winds, \citet{BaiStone13b}
analyzed the local wind kinematics and concluded that wind
acceleration is centrifugal. This apparent discrepancy is
due to the different reference frames adopted.
The local shearing box is formulated in a frame corotating
with the disk, where the force equation becomes [see eq.~\eqref{eq:eng} \& \citealp{Spruit96}]
\begin{equation}
\frac{dv'^2}{ds}=-\frac{d}{ds}\bigg(h+\Phi_{\rm eff}\bigg)\ .
\end{equation}
Here, $v'$ is the total velocity in a frame corotating with the field line (angular
velocity $\omega$), where its acceleration is mainly compensated
by the drop in $\Phi_{\rm eff}$. However, when corotation is not
enforced, the change in $v'$ mainly reflects the deviation of
$v_\phi$ from $\omega R$ rather than acceleration of $v_p$.
Indeed \citet{BaiStone13b} pointed out that the wind
launching process is largely due to the vertical magnetic pressure
gradient of the toroidal field. Our results here show that this 
continues to be true of a global wind.

The fact that typical PPD winds are expected to be launched by magnetic
pressure gradients raises concerns about stability. With toroidal field
highly dominating poloidal field from the wind base, the wind velocity below the
Alfv\'en point $v_p=v_{Ap}$ is already much smaller than the total Alfv\'en
velocity. For a single flux rope, kink instability is expected to develop on
Alfv\'en-crossing timescales, although the situation is more complicated
and less clear for disk winds with extended flux distributions \citep{Spruit96}.
This question needs to be investigated via 3D MHD simulations. Previous
2D axisymmetric wind simulations also reported that no stable wind
solution can be found for sufficiently low magnetization (e.g.,
\citealp{Murphy_etal10,Sheikhnezami_etal12}), although higher resolution
and more realistic physical input are needed to clarify the threshold
magnetization.
Magnetic pressure driven outflows have also been identified
in star formation simulations \citep{Machida_etal08,HennebelleFromang08},
and the kink instability appears to disturb the coherent toroidal field structure
without affecting the acceleration \citep{Tomida_etal13}. 
On the other hand, recent 3D disk wind simulations by \citet{SheikhnezamiFendt15}
with relatively strong magnetization found no sign of kink instability.

\begin{figure}
    \centering
    \includegraphics[width=85mm,height=180mm]{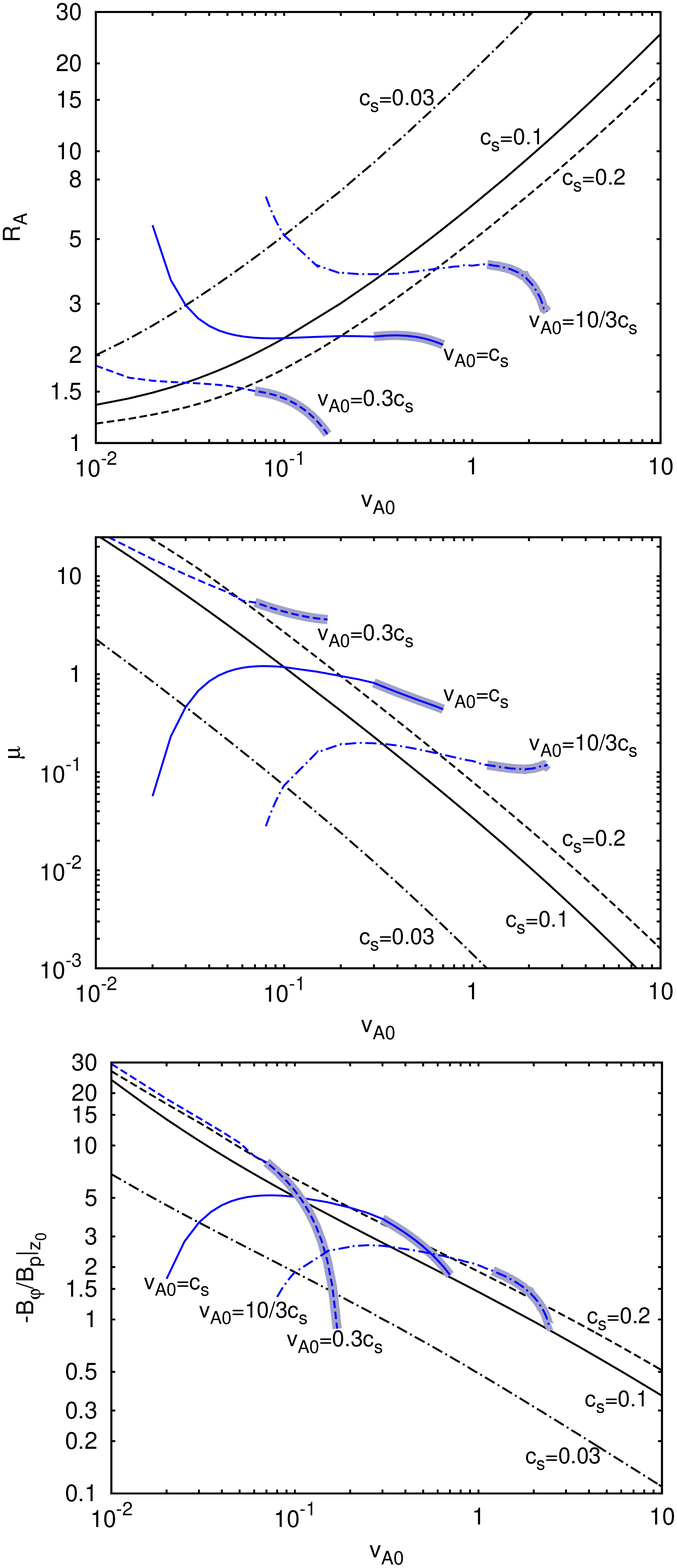}
  \caption{Dependence of Alfv\'en radius $R_A$ \emph{(top)}, mass loading
  $\mu$ \emph{(middle)} and $|B_\phi/B_p|_{\mathrm{base}}$ \emph{(bottom)} on $v_{A0}$ for
  different choices of $\csw$ (fixed or $\propto v_A$), as marked, with
  $q=0.25$ throughout.
  Shaded regions of blue lines have $R_s<R_0$, which
  violates our assumptions.}\label{fig:thermal}
\end{figure}

\subsection[]{Role of Wind Thermodynamics}\label{ssec:temp}

\begin{figure*}
    \centering
    \includegraphics[width=180mm]{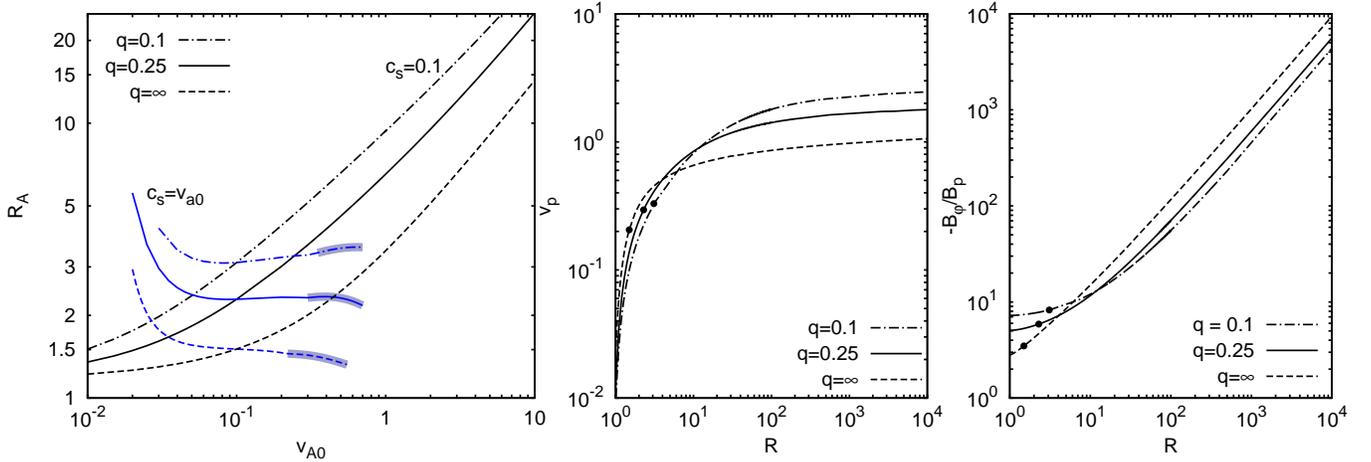}
    \caption{\emph{Left:} Alfv\'en radius $R_A$ as a function of $v_{A0}$ for different
      field divergences $q$  [eq.\eqref{eq:bpR}] as marked in the legend. Black
      lines for fixed $\csw=0.1$, blue lines for $\csw=v_{A0}$.  Shaded
      portions have $R_s<R_0$,  violating our assumptions.  \emph{Middle:} poloidal
      wind velocity along field lines for solutions with different $q$, as
      marked, and fixed $\csw=v_{A0}=0.1$.  Black dots mark Alfv\'en points.
      \emph{Right:} similar to the middle panel, but for the field pitch
      $-B_{\phi}/B_p$.}\label{fig:divergence}
\end{figure*}

The balance among heating, radiative cooling, and adiabatic expansion
affects the wind properties in two ways.
First, it affects the internal wind temperature, which we represent
by a constant $\csw$ (isothermal). We will further explore
different equation of states in \S\ref{ssec:eos} and show
that the additional effect is minor. Second, it affects the location
of the wind base $z_0$. As discussed in Appendix \ref{app:bstr},
at fixed field strength, $v_{A0}$ depends exponentially on $z_0$:
deeper penetration would reduce $v_{A0}$.
While we fix $z_0=0.15R_0$ in this section, we show in \S
\ref{ssec:geom} that the absolute value of $z_0$ is unimportant,
thus the penetration depth affects the wind properties mainly
by affecting $v_{A0}$.

In the top panel of Figure \ref{fig:thermal}, we fix $q=0.25$, and show the dependence of
Alfv\'en radius on $v_{A0}$ for two different values of $\csw$. According to our
discussions above, black lines (fixed $\csw$, varying $v_{A0}$) can be considered as
varying the penetration depth at fixed wind temperature and field strength.  Some
consequences have been discussed in the previous subsection, but
are now endowed with a new interpretation. Namely, enhanced penetration of external
radiation would reduce the importance of magnetic fields, leading to a weaker wind with
lower terminal velocities, smaller Alfv\'en radius, and more toroidally dominated (tightly
wound) magnetic field.

The blue lines in the left panel correspond to varying the wind
temperature at fixed penetration and field strength (see Section
\ref{ssec:units}). We see that over a large range, these lines are
relatively flat, suggesting that the location of $R_A$ is to a large
extent controlled by the ratio of $v_{A0}/\csw$ at the wind launching
region. The contrast among these lines is very significant. Varying
$v_{A0}/c_s$ around unity by a factor of $\sim3.3$
makes the $R_A/R_0$ vary from $\sim2.3$ to $\sim1.4$
and $\sim4$, Because the ratio $\dot{M}_w/\dot{M}_{a}$
depends quadratically on $R_A/R_0$, such variations in the latter would easily
give variations in wind mass loss rate by factors $\sim10$.
This discussion best demonstrates the sensitive dependence of the
wind mass loss rate on the thermodynamics.

In reality, stronger external radiation is likely to lead to both higher
wind temperature and deeper penetration. Therefore, enhancing
external radiation may correspond to moving in between rightward
(larger $\csw$) and to the bottom left (deeper penetration) in the top
panel of Figure \ref{fig:thermal}, along directions between
tangents to the blue and black curves.

The middle and bottom panels of Figure \ref{fig:thermal} further show
the dependence of the mass loading parameter $\mu$ and
toroidal-to-poloidal field ratio at the wind base $|B_\phi/B_p|_{z_0}$.
They are directly relevant to calculate the wind-driven accretion and
wind mass loss rate via eqs.~(\ref{eq:diracc}) \&
(\ref{eq:dirmloss}).

Similar to the discussions above, increasing the penetration depth
at fixed temperature (black lines, towards smaller $v_{A0}$) yields both
larger mass loading, and larger $|B_\phi/B_p|$ at wind base. The increase
in $\mu$ is in fact faster, leading to higher ratio of mass loss rate over
wind-driven accretion rate, in accordance with the trend seen for $R_A$ in
the top panel. Note again that this discussion applies equally to varying
poloidal field strength at fixed temperature. Increasing the temperature
at fixed poloidal field strength (blue lines, towards larger $v_{A0}$) yields
similar trend in both $\mu$ and $|B_\phi/B_p|_{z_0}$ (ignoring the
shaded portion of the lines), consistent with the top panel where the blue
lines are largely horizontal over a wide range of $v_{A0}$.

\begin{figure*}
    \centering
    \includegraphics[width=160mm]{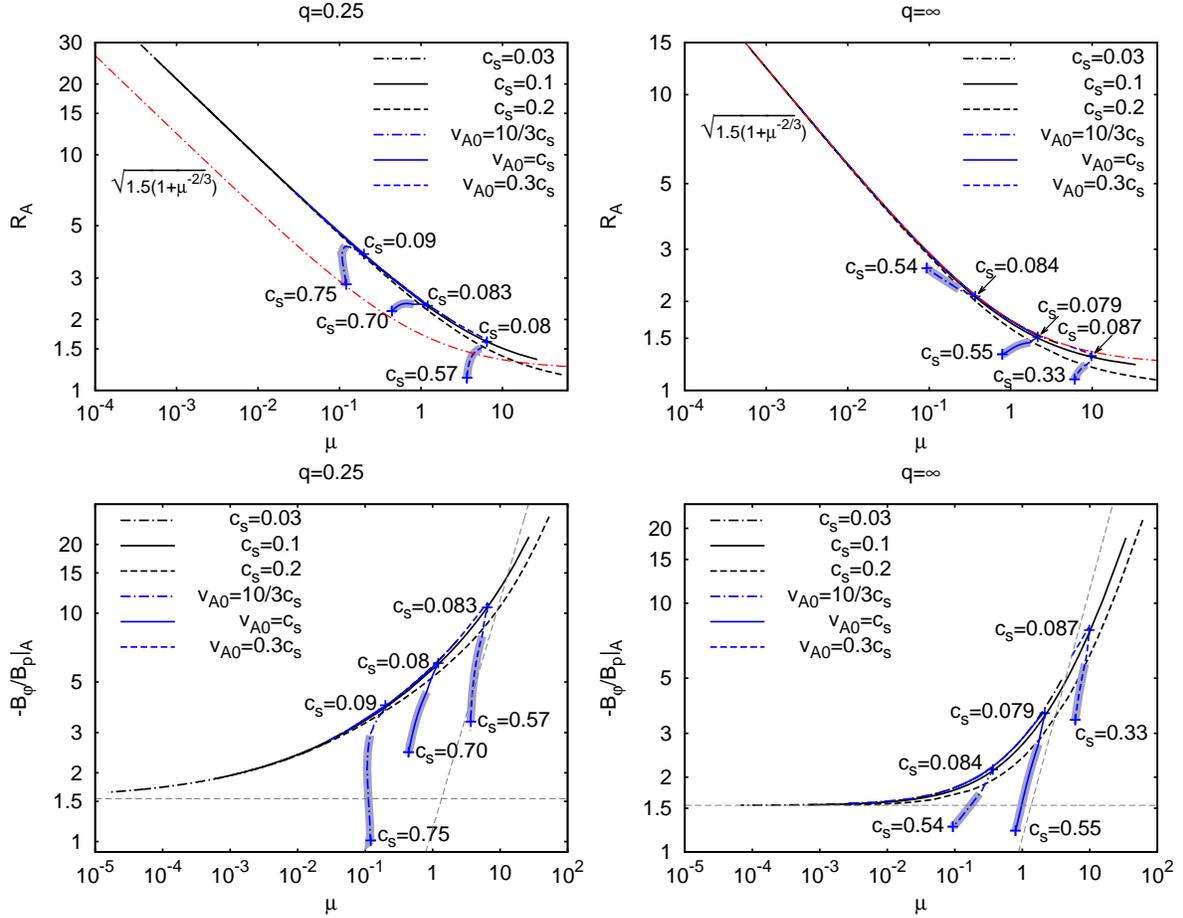}
  \caption{Scaling relations between pairs of wind diagnostics.
  The top two panels show the relation between the mass loading
  parameter $\mu$ and $R_A$, while the bottom two panels show
  the scaling relation between $\mu$ and $(-B_\phi/B_p)|_{R_A}$.
  Left (right) panels correspond to wind solutions with $q=0.25$
  ($q=\infty$). In each panel, we either choose fixed $\csw$, or
  choose $\csw$ to be proportional to $v_{A0}$, as marked in the
  legends. To guide the eye, we further label the value of $\csw$
  at various turning and tip points in the blue lines. The red dash-dotted
  line in the top panels indicates the expected scaling relation
  for the cold Weber \& Davis wind model. Similarly, the gray dashed
  lines in the bottom panels are the expected asymptotic scaling
  relations. Shaded regions have solutions with $R_s<R_0$, which would
  violate our assumptions, but are still shown for reference.}\label{fig:scaling1}
\end{figure*} 

\subsection[]{Role of Field Line Divergence}

The field line divergence parameter $q$ characterizes the location
$R_t\approx q^{-1/2}R_0$ where field line geometry transitions from being approximately
parallel to diverging. Besides the
fiducial value $q=0.25$, we explore two additional
cases with $q=0.1$ and $q=\infty$.  The main results are shown
in Figure \ref{fig:divergence}.

One sees that the Alfv\'en radius increases as $q$ decreases.  Note that at fixed $B_{p0}$
(or $v_{A0}$), smaller $q$ yields a stronger $B_p$ at $R>R_0$.  Therefore, the
trend of $R_A$ with $q$ is consistent with expectations from \S \ref{ssec:Bpstr}.  In
addition to larger $R_A$, smaller $q$ values yield higher terminal speeds, and less
tightly wound fields, as shown in the middle and right panels. It is also interesting to
note that the wind field is more toroidally dominated at the wind base for smaller (rather
than larger) $q$, but the trend reverses at larger $R$.   Therefore, smaller $q$ makes
the magnetic pressure gradient more dominant in accelerating the wind.

Given the field strength and wind temperature, a lower limit of $R_A$
is set by $q=\infty$, where field lines diverge from the beginning.
For $v_{A0}=\csw=0.1v_K$, we obtain $R_A\approx1.51$, as
opposed to $R_A=2.29$ when $q=0.25$. Our choice of $q=0.1$
corresponds to $R_t\sim3R_0$, which is already sufficiently far from
the wind launching region.\footnote{In the unphysical situation of
$q\rightarrow0$ (hence $B_p\propto 1/R$), no wind solution can be
found.} Therefore, this case
may be considered to set an upper limit to $R_A$, giving
$R_A\approx3.11$ under the same conditions.

Overall, we see that even at fixed $v_{A0}$ and $\csw$, the possible
values of $R_A$ span a wide range depending on how rapidly poloidal
field lines diverge, leading to variations in
$\xi=(d\log{\dot{M}_w}/d\log{R})/\dot{M}_a$ by a factor of $\sim7$.
Reading figures from many existing wind simulations with different
setups, it appears that the steady-state poloidal field configurations tend
to diverge well before reaching $R\sim3R_0$ (e.g.,
\citealp{Krasnopolsky_etal99,Zanni_etal07,Murphy_etal10,StepanovsFendt14}).
In reality, the situation must depend on the radial distribution of magnetic flux.
For instance, if magnetic flux were strongly concentrated towards the inner disk,
then one would expect a $B_p\propto R^{-2}$ scaling right from the wind
launching region, as can be seen in \citet{Krasnopolsky_etal03}. The
X-wind model \citep{Shu_etal94} constitutes an extreme case.
The relatively large ejection efficiency $\xi$ in these models is likely
(partially) compensated by a relatively narrow radial range in which the wind is launched,
thus limiting the total wind mass loss rate. On the other hand, a
quasi-self-similar field configuration (i.e., small $q$) would yield much smaller
$\xi$, but with a much more extended radial range. Thus, the contrast in total
wind mass loss rate between large and small $q$ cases may not be as
dramatic as the difference in $\xi$.

\begin{figure*}
    \centering
    \includegraphics[width=160mm]{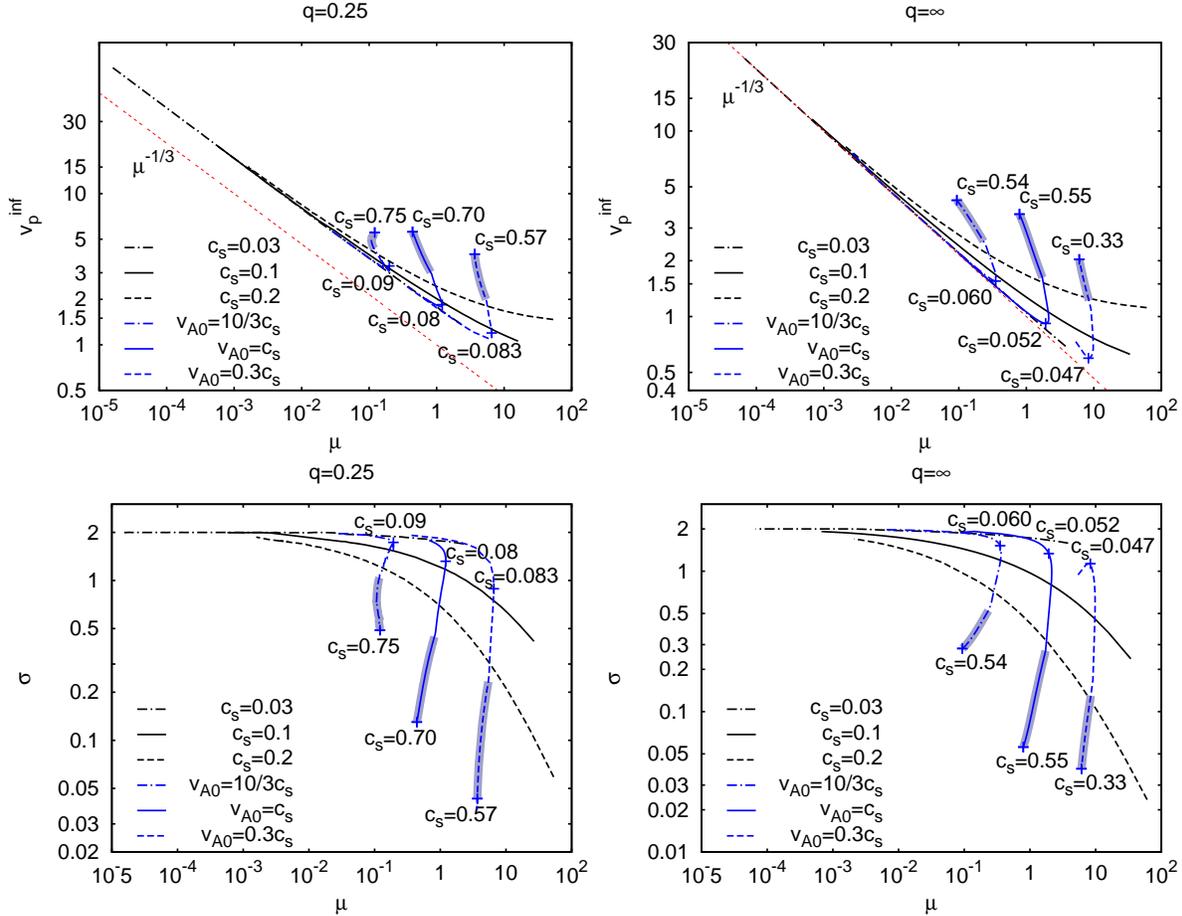}
  \caption{Scaling relations between pairs of wind diagnostics.
  The top two panels show the scaling relation between the mass loading
  parameter $\mu$ and the asymptotic poloidal velocity $v_p^{\rm inf}$,
  while the bottom two panels show the relation between $\mu$ and
  $\sigma$ (asymptotic ratio of Poynting to kinetic energy flux).
  Left (right) panels correspond to wind solutions with $q=0.25$
  ($q=\infty$). In each panel, we either choose fixed $c_s$, or
  choose $\csw$ to be proportional to $v_{A0}$, as marked in the
  legends. The red dash-dotted line in the top panels indicates the
  expected scaling relation for the cold Weber \& Davis wind model
  (\ref{eq:muvp}).  To guide the eye, we further label the value of
  $\csw$ at various turning and tip points. Shaded regions have
  solutions with $R_s<R_0$, which would violate our assumptions, but
  are still shown for reference.}\label{fig:scaling2}
\end{figure*}

\subsection[]{Scaling Relations}\label{ssec:scaling}

In \S\ref{ssec:diag}, we provided several scaling relations among
various characteristic wind quantities that are expected in the Weber \&
Davis wind ($\theta=0$, $q=\infty$, $\csw\rightarrow0$).   Our models
being generalizations of WD's, we test in this subsection to what extent they
follow the same scalings.

In Figure \ref{fig:scaling1}, we test the scaling relations (\ref{eq:mura}) and
(\ref{eq:mubphi}), namely, the dependence of $R_A$ and $|B_\phi/B_p|_{R_A}$ on the mass
loading parameter $\mu$ [eq.~\eqref{eq:mu}]. These two relations state that large mass
loading is accompanied by small $R_A$, and more toroidally dominated field configuration,
which means that the wind is more efficient at removing mass than angular momentum.

We see that for the $q=\infty$ cases, the scaling
relation (\ref{eq:mura}) is matched almost perfectly as long as the wind
is sufficiently cold $\csw\lesssim0.08v_K$.  This relation seems to be
universal regardless of field inclination. It is interesting to note that
even when $v_{A0}<\csw$, the relation still holds as long as
$\csw\lesssim0.08v_K$.  The relation has also been
approximately recovered in global wind simulations \citep{Zanni_etal07}.
For hotter winds with $\csw\gtrsim0.08v_K$, $R_A$ falls below what is
expected from the relation at a given $\mu$. In particular, the minimum
value of $R_A$ is $1.5$ for a cold wind according to eq.~\eqref{eq:mura}, but
this is no longer a lower limit if the wind is warm or hot.

In the case that $q=0.25$, one sees that while for a cold wind with
$\csw\lesssim0.08v_K$, all data points line up into an almost perfect
scaling relation, it is clearly offset from eq.~\eqref{eq:mura}. As can be partly
traced from Figure \ref{fig:divergence}, for fixed $v_{A0}$ and $\csw$
the $q=0.25$ case gives both larger $\mu\propto\rho v_p$ and larger
$R_A$ compared to the $q=\infty$ case, thus shifting the original
relation towards the upper right, as is observed here. Other aspects of the
relation are very similar to the $q=\infty$ case.

\begin{figure*}
    \centering
    \includegraphics[width=180mm]{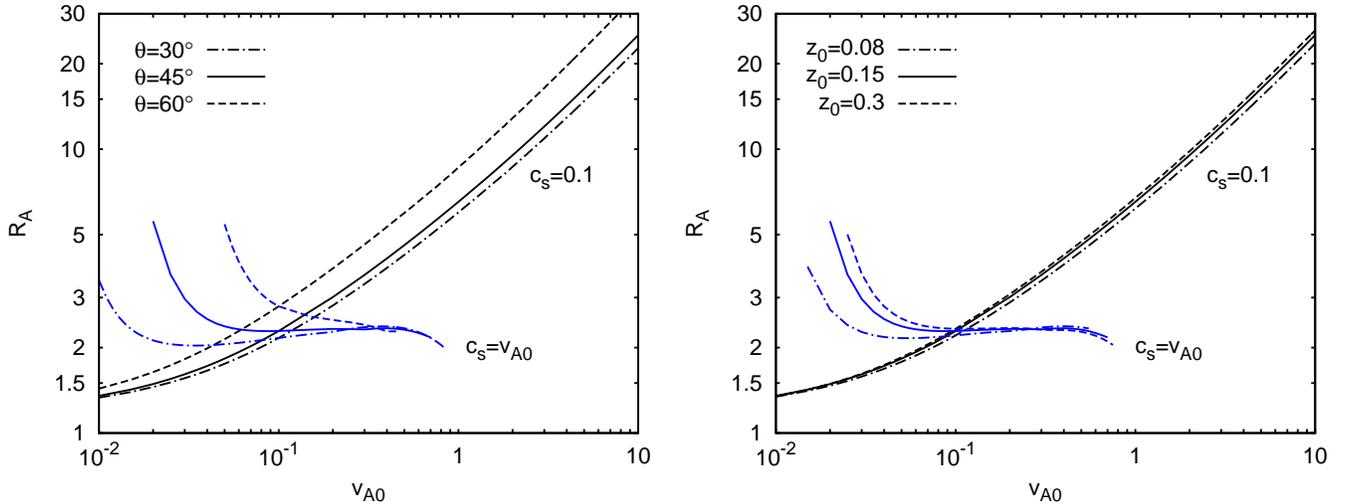}
  \caption{Left: Alfv\'en radius $R_A$ as a function of $v_{A0}$ for
  different choices of field inclination angle $\theta$, as marked in
  the legend. Black lines have fixed $\csw=0.1$, while blue lines
  have fixed $\csw=v_{A0}$. Other parameters are taken as
  fiducial (e.g., $q=0.25$, $z_0=0.15$). Right: same as left, but
  for different choices of wind base location $z_0$. Other parameters
  are taken as fiducial (e.g., $q=0.25$, $\theta=45^\circ$).}\label{fig:geom}
\end{figure*}

In the bottom panels of Figure \ref{fig:scaling1}, one sees that when
$q=\infty$, the asymptotic relation (\ref{eq:mubphi}) is accurately satisfied
in the strong-field limit (giving $\mu\ll1$), with
$|B_\phi/B_p|_A\rightarrow1.5$. The weak field limit of the asymptotic
relation depends on $\csw$, and is only achieved with
$\csw\rightarrow0$. In the case of $q=0.25$, there is also a
well-defined relation for $|B_\phi/B_p|_A$ in the strong field limit,
although it does not appear to approach a constant value. Similarly,
the weak field limit of the relation depends on wind temperature.
One general trend is that higher wind temperatures lead to smaller
$|B_\phi/B_p|_A$. The main reason is that higher temperature
provides stronger pressure support that induces sub-Keplerian rotation,
which would reduce the development of toroidal fields.

In Figure \ref{fig:scaling2}, we further test the scaling relation
(\ref{eq:muvp}) in the top panels. We see that the $\mu^{-1/3}$
scaling for terminal velocity $v_p^{\rm inf}$ is very well satisfied
when $q=\infty$ if  the wind is cold. Enlarging the temperature
at fixed $v_{A0}$ yields larger $v_p^{\rm inf}$. When $q=0.25$,
our fiducial choice,  the $\mu^{1/3}$ scaling is still
very well satisfied, except for being shifted up by a factor of $\lesssim2$,
consistent with Figure \ref{fig:divergence}.

The asymptotic ratio of Poynting flux to kinetic energy ($\sigma$) for
various wind parameters is shown in the bottom panel of Figure
\ref{fig:scaling2}.  Regardless of the $q$ value, $\sigma$
always reaches the maximum value of $2$ as long as the wind is
strong (giving small $\mu$), or cold (small $\csw$ and
$v_{A0}<\csw$).  Modestly warm winds with
$\csw\gtrsim0.1v_K$ and solutions with $v_{A0}<\csw$
all yield smaller Poynting flux, and the wind power is then largely kinetic.

\section[]{Parameter Dependence}\label{sec:param}

In this section, we explore the secondary parameters, whose effect
on wind properties turns out usually to be minor,
justifying in part the simplicity of our models.

\subsection[]{Dependence on Field Geometry}\label{ssec:geom}

We first explore the effect of the field inclination angle $\theta$ and
wind-base height $z_0$. Since we are mostly interested in the efficiency
of angular momentum transport and mass loss, we only show the
dependence of $R_A$ on $v_{A0}$ for various values of $\theta$ and
$z_0$, illustrated in Figure \ref{fig:geom}.

The inclination angle influences the effective potential that the wind
flow experiences. Larger angles require the wind flow to climb
out of a deeper gravitational potential well. As a result, hydrostatic
support extends farther and the density at the slow critical point drops, leading to
a more lightly loaded wind and more effective acceleration.
This trend is clearly observed in Figure \ref{fig:geom}. Increasing (decreasing)
$\theta$ always leads to larger (smaller) $R_A$ at fixed wind temperature.
The change in $R_A$ is more pronounced when varying $\theta$ from
$45^\circ$ to $60^\circ$ than from $30^\circ$ to $45^\circ$.

When fixing $\csw=v_{A0}$, we see that the effect of different inclination
angle is more significant towards lower temperature. This is simply because
higher temperature helps overcome the steeper gravitational potential
caused by larger $\theta$, and in general reduces the importance of gravity.
For our fiducial choice $v_{A0}=\csw=0.1v_K$,  appropriate for
$R\sim1\,\au$ in PPDs, the difference in $R_A$ among different
choices of $\theta$ is modest unless the inclination angle is
larger than $60^\circ$. Furthermore, at larger radii in PPDs, we expect
$\csw/v_K$ to increase, thus making the inclination effect only a minor
consideration.

The right panel of Figure \ref{fig:geom} shows the corresponding plot for varying
$z_0$. Clearly, the geometric location of the wind base has very little impact on the wind
properties, apart from a very weak tendency to be more lightly loaded with
increasing $z_0$. Note that here we simply vary the geometric location of the wind base
$z_0/R_0$, but we do not vary $v_{A0}$ according to (\ref{eq:va0par}). The latter effect
has already been incorporated in the discussions in Section \ref{ssec:temp}.  Evidently,
the UV penetration depth affects the wind properties mainly by
controlling the Alfv\'en speed $v_{A0}$ rather than the geometric height $z_0$.

In BP and subsequent literature, it is often stated that $60^\circ$ is the critical angle for
launching a magneto-centrifugal wind. However, this statement applies to
cold winds launched from razor-thin disks
($\csw=0$ and $z_0=0$).
For winds of finite temperature launched from finite heights,
there is no sharp inclination threshold.

One main limitation of this work is that we have prescribed the poloidal field geometry
rather than solving for it self-consistently from cross-field force balance.  The main
effect that we miss is wind collimation, which is a natural consequence of disk wind from
magnetic hoop stress, and is present in almost all self-similar wind solutions and wind
simulations. The fact that our wind solutions are not very sensitive to variations in
$\theta$ within $30^\circ-60^\circ$ already suggests that our main results are not
sensitive to field geometry. In addition, significant collimation is typically achieved
well beyond the Alfv\'en point, where the wind flow has largely escaped from the
gravitational potential.  Thus we expect essential wind properties (viz. mass loading and
$R_A/R_0$) not to be strongly affected by the field geometry at large distances.

\begin{figure}
    \centering
    \includegraphics[width=93mm]{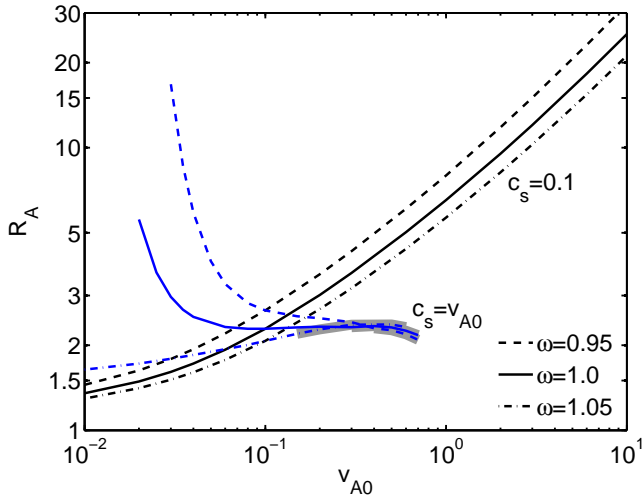}
    \caption{Alfv\'en radius $R_A$ versus Alfv\'en speed at wind base $v_{A0}$ for
      different field-line angular velocities $\omega$, as marked in the legend. Black
      lines have fixed $\csw=0.1$, blue lines $\csw=v_{A0}$.  Other parameters are
      fiducial (e.g., $q=0.25$, $\theta=45^{\circ}$). Shaded regions have $R_s<R_0$, which
      violates our assumptions.  }\label{fig:omega}
\end{figure}

\subsection[]{Dependence on Magnetic Angular Velocity}\label{ssec:omega}

Heretofore we have fixed $\omega=\Omega_K$,
which is expected to hold approximately but not strictly. A
source of deviation of $\omega$ from $\Omega_K$ is intrinsic to the
disk rotation profile in PPDs, which are externally irradiated and flared.
The midplane rotates at sub-Keplerian speed due to partial radial pressure
support. On the other hand,
super-Keplerian rotation may occur at the disk surface where the
radial pressure gradient reverses.   For the standard vertically
isothermal MMSN model (see Appendix \ref{app:bstr} for adopted
parameters), the non-Keplerian deviation is
\begin{equation}
\frac{\Delta v_\phi}{v_K}\approx1.1\times10^{-3}\bigg(\frac{13}{8}R_{\au}^{1/2}
-\frac{5}{8}\frac{z^2}{H^2}R_{\au}^{-1/2}\bigg)\ ,
\end{equation}
where $R_{\au}$ is radius measured in $\au$. For our fiducial choice
of $R=1\,\au$ and $z_0=4.5H$, we get super-Keplerian rotation by about
$1.2\%$.  At larger radii, we find sub-Keplerian rotation
$<2\%$ out to $\sim120\,\au$.

Given these estimates, we show in Figure \ref{fig:omega} the Alfv\'en
radius as a function of $v_{A0}$ for $\omega$ varying
at an exaggerated $5\%$ level (all other parameters being
fiducial). We see that at fixed $\csw=0.1$, varying $\omega$ at this level
has relatively small impact on the wind properties.  Thus
we do not expect wind properties to be sensitive to $\omega$. 

One sees that smaller $\omega$ leads to larger $R_A$ 
(except for hotter winds as marked in shaded areas where
$R_s<R_0$ and our solutions are less trustworthy). This is because smaller
$\omega$ is associated with less
free energy available for wind launching, requiring
a larger density drop in the hydrostatic region near the wind base, and
leading to a more lightly loaded wind. If sub-Keplerian rotation is
substantial, wind launching can be suppressed without thermal pressure
support \citep{Shu_etal08}. In Figure \ref{fig:omega}, the blue
lines diverge towards smaller $\csw$
(chosen equal to $v_{A0}$), while tending to converge towards
large $\csw$. This reflects the importance of thermal effects:
colder winds are strongly affected by the rotation profile at the wind base
due to energetic constraints, while warmer winds are exempt from such
sensitive dependence. Therefore, external heating of PPD surface
eases the wind launching process and makes it suffer much less from
deviations from Keplerian rotation.

\begin{figure}
    \centering
    \includegraphics[width=90mm]{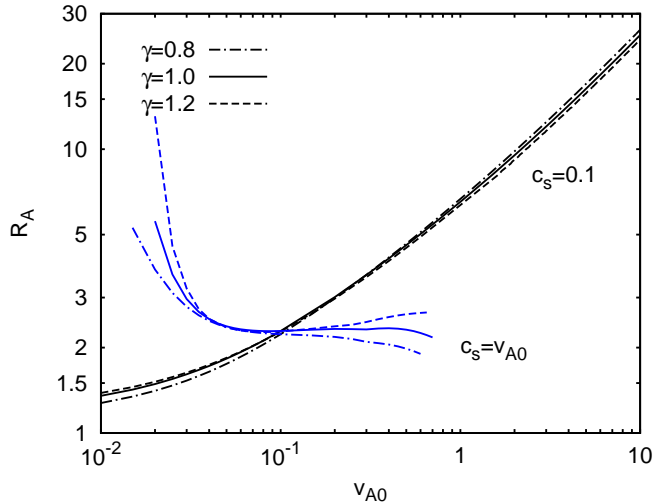}
    \caption{Like Fig.~\ref{fig:omega} but for differing polytropic indices $\gamma$
    rather than angular velocities.}\label{fig:eos}
\end{figure}

\subsection[]{Dependence on Equation of State}\label{ssec:eos}

Our discussion of the role of thermodynamics has been restricted
to an isothermal equation of state. To justify this
choice, here we  substitute a polytropic equation of state with
$P=K\rho^\gamma$. This change enters the Bernoulli
equation through the enthalpy term, which now becomes
\begin{equation}
h=\frac{\gamma}{\gamma-1}K\rho^{\gamma-1}\ .
\end{equation}
We set the parameter $K$ such that $P/\rho=\csw^2$ at the wind base. We consider two
adiabatic indices: $\gamma=0.8$ and $1.2$. The former mimics the increase of gas
temperature above the wind base as a result of additional heating sources that penetrate
no less deeply than FUV (e.g., extreme UV). The latter ($\gamma=1.2$) leans toward the
opposite case where adiabatic cooling reduces gas temperature along wind field lines.

Figure \ref{fig:eos} illustrates the results of these different equations of
state.   Evidently the Alfv\'en radius is very insensitive to
variations in $\gamma$. We have also checked the mass loading
parameter $\mu$ and $|B_\phi/B_p|$ at the wind base, and confirm that
they are all very similar for different choices of $\gamma$s. Looking
at Figure \ref{fig:contour}, the density drop from the wind base (near
the slow point) to the Alfv\'en point in our fiducial solution amounts to
more than 2 orders of magnitude. Such density drop leads to
temperature variations for our different choices of $\gamma$s by a
factor more than $6$, yet major wind properties remain similar to
the isothermal case.


Different choices of $\gamma$ do affect a few other wind properties.
In particular,  for the somewhat unphysical choice
$\gamma=0.8$, the gas temperature increases with decreasing
density. As a result, the gas never reaches a terminal speed but keeps
accelerating on its own: magnetic field is not required for wind launching,
a situation analogous to pure photo-evaporation. The $\gamma=1.2$
case, on the other hand, shows very little difference in terminal
velocity from the isothermal one. Also, the location of fast
magnetosonic point is extremely sensitive to $\gamma$:
$R_f=64.2, 554.5$ and $7.70\times10^4$ for $\gamma=0.8$, $1$ and
$1.2$, respectively.
Despite such differences, we have already seen that our most interesting
quantity, the Alfv\'en radius, and the associated wind-driven accretion rate
and wind mass loss rate, are very insensitive to thermodynamics at $R\gg R_0$.

\section[]{Discussion}\label{sec:discussion}

\subsection[]{Observations of PPD Outflows}\label{ssec:obs}

Disk winds from PPDs likely correspond to the low-velocity component of outflow (as
opposed to the high-velocity jet), whose signatures have now been routinely inferred from
blue-shifted emission line profiles such as from CO, OI and NeII lines (e.g.,
\citealp{Hartigan_etal95,Pascucci_etal09,Pontoppidan_etal11,Bast_etal11,
  Herczeg_etal11,Sacco_etal12,Rigliaco_etal13,Brown_etal13,Natta_etal14}). The blueshift
is typically a few km s$^{-1}$, comparable to the disk sound speed at sub-$\au$ to $\au$
scales, and is consistent with MHD wind flows near the launching region.

One crucial parameter in the wind-driven accretion framework is the ratio of wind
mass loss rate $\dot{M}_{\rm wind}$ to the wind-driven accretion rate
$\dot{M}_{\rm acc}$.  A correlation between the two quantities was
identified in early observational studies \citep{Cabrit_etal90,Hartigan_etal95}.
Following later calibration of accretion rates \citep{Gullbring_etal98,Muzerolle_etal98}, a canonical
number of $\dot{M}_{\rm wind}/\dot{M}_{\rm acc}\approx0.1$ has been established,
i.e., mass loss rate is a small fraction of accretion rate. However, the mass-outflow rate
obtained by \citet{Hartigan_etal95} corresponds to the high-velocity 
outflow, which most likely originates from a jet launched close to the protostar.
While low-velocity disk winds were ubiquitously detected in \citet{Hartigan_etal95},
reliable estimates of the mass-flow rate were not feasible. Recently, \citet{Natta_etal14}
modeled the low-velocity component of the blueshifted forbidden lines from a
sample of T-Tauri stars, and estimated $\dot{M}_{\rm wind}/\dot{M}_{\rm acc}\approx0.1-1$.
Estimates of mass outflow rates from more embedded protostars by \citet{Watson_etal15}
also suggest the ratio of $\dot{M}_{\rm wind}/\dot{M}_{\rm acc}$ in the similar range.
Studies by \citet{Klaassen_etal13} and \citet{Salyk_etal14} for two
individual sources (HD~163296 and AS~205N) using the Atacama Large Millimeter
Array (ALMA) both indicate significant mass outflow and again suggest
$\dot{M}_{\rm wind}/\dot{M}_{\rm acc}\approx0.1-1$ (with large uncertainties).

Overall, current observational constraints, although limited and bearing large
uncertainties, agree with our suggested scenario that PPDs lose mass at a
considerable fraction of the accretion rate. In the future, it will be crucial to combine
observational and modeling efforts to reduce the uncertainties and better
constrain the physical nature of PPD disk winds.

\begin{figure*}
    \centering
    \includegraphics[width=180mm]{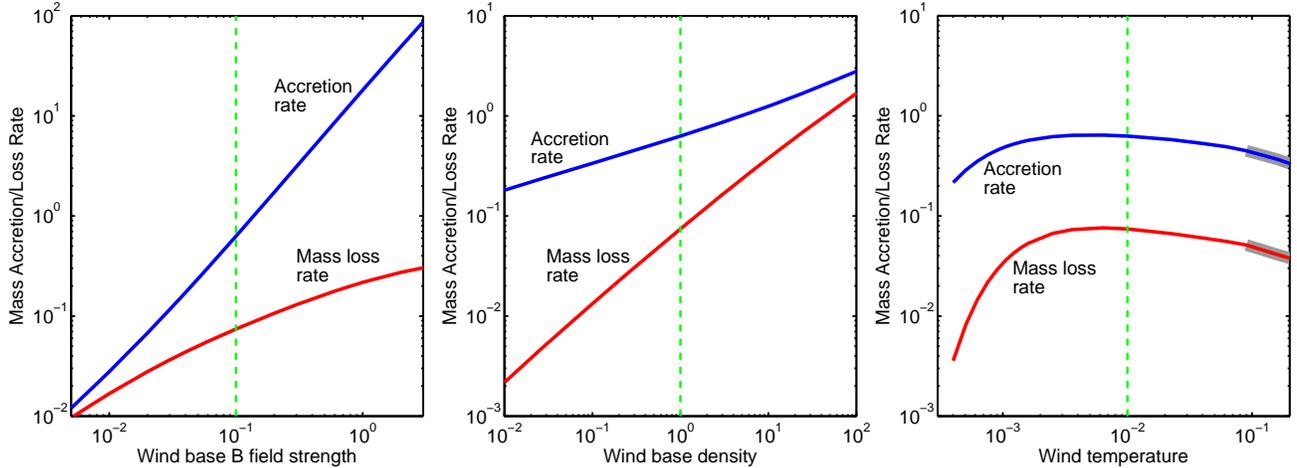}
    \caption{Schematic dependence of wind-driven accretion rate and mass-loss rate on the
      main physical parameters of magneto-thermal winds, constructed based on fiducial
      wind model parameters shown in Fig.~\ref{fig:thermal} (choosing $\csw=0.1v_K$ for
      left and middle panels and $\csw=v_{A0}$ for the right panel) and
      eqs.~\eqref{eq:diracc} and \eqref{eq:dirmloss}. Our standard wind solution
      ($v_{A0}=\csw=0.1v_K$) is marked by green dashed lines, corresponding to
      an accretion rate $\sim10^{-8}M_{\bigodot}$ yr$^{-1}$ at 1~$\au$.  The main
      parameters are poloidal magnetic field strength at the wind base $B_{p0}$ (\emph{left}, in
      units of $v_{A0}$), density of the wind base $\rho_b$ (\emph{middle}, reflecting the
      penetration depth of FUV ionization, in arbitrary units), and wind temperature $T_w$
      (\emph{right}, in units of $\csw^2/v_K^2$).}\label{fig:maccmloss}
\end{figure*}

\subsection[]{Magneto-photoevaporation?}\label{ssec:magpe}

As mentioned in the introduction, mass loss from PPDs has conventionally
been attributed to photoevaporation. This is a purely hydrodynamic
process whereby the disk surface is heated by external radiation to sufficiently
high temperatures to launch a thermal wind.
Early models of photoevaporation considered heating from extreme UV
(EUV, \citealp{Hollenbach_etal94,Font_etal04,Alexander_etal06a}). These
calculations gave mass-loss rates of the order of
$10^{-10}-10^{-9}M_{\bigodot}$ yr$^{-1}$, and when coupled with a viscously
evolving disk, can lead to rapid disk dispersal 
\citep{Clarke_etal01,Alexander_etal06b}. Later theoretical calculations
included heating from X-rays
\citep{Alexander_etal04,Ercolano_etal08,Ercolano_etal09,Owen_etal10}
and far-UV \citep{Adams_etal04,GortiHollenbach09,Gorti_etal09,Gorti_etal15}
with increasing levels of complexity. These works demonstrated that due to
deeper penetration, X-ray and FUV can drive stronger winds and
mass-loss rates $\sim 10^{-9}-10^{-8}M_{\bigodot}$ yr$^{-1}$,
which approach typical disk accretion rates.

There are two major differences between the photoevaporation
and MHD-disk-wind scenarios. First, a thermal wind exerts no torque
on the disk, and hence the mass-loss rate is unrelated to the accretion
process. Second, magnetic forces plays a dominant role in wind driving in
an MHD wind (see Figure \ref{fig:force}), and with the additional push, likely
yield larger mass loss rate. Unfortunately, it may not be easy to
distinguish the two scenarios observationally. Along flow streamlines,
specific angular momentum is conserved for a purely thermal wind. In MHD
winds, unless the poloidal field is strong and the wind lightly loaded,
specific angular momentum is also approximately conserved, as seen
in Fig.~\ref{fig:prof}.

In reality, it is probably meaningless to argue for which scenario is more
correct. It is clear that both external irradiation
and magnetic fields play indispensable roles in the wind launching and
acceleration processes. In this regard, we feel that  ``magneto-thermal
wind'' or ``magneto-photoevaporation" may be more appropriate names to
describe the mass loss process from PPDs.

\subsection[]{Global Evolution of Protoplanetary Disks}

Our simple 1D wind model represents a starting point towards constructing a global
model of PPD evolution that predicts the rate of angular momentum transport and mass
loss at all disk radii. 

We first note that while our model is mainly motivated from studies of PPD gas dynamics in
the inner disk, our wind framework is likely equally applicable to the outer disk as well
($R\gtrsim10$ AU). At the outer disk, ambipolar diffusion is the sole dominant non-ideal
MHD effect, which substantially reduces the coupling between gas and magnetic fields in the
disk interior, and damps or suppresses the MRI \citep{BaiStone11,Simon_etal13a,Simon_etal13b}.
Efficient angular momentum transport requires the presence of net vertical magnetic flux, which
in the mean time launches a disk wind. Far UV ionization again has been found to be crucial
in these processes, and if penetrating sufficiently deeply, can lead to vigorous MRI turbulence
at disk surface \citep{Simon_etal13b,Bai15}. On the other hand, non-detection of surface
turbulence around the disk HD 163296 by \citet{Flaherty_etal15} may suggest less deep
FUV penetration, leading to a largely laminar disk upon which our wind framework can be readily
applied.

To facilitate an effort towards constructing a global model of wind-driven PPD evolution,
we show in Figure \ref{fig:maccmloss} the dependence of wind-driven
accretion rate $\dot{M}_{\rm acc}$,  and wind mass loss rate per logarithmic radius
$d\dot{M}_{\rm wind}/d\ln R$, as a function of the main wind parameters, based
on Fig.~\ref{fig:thermal} and eqs.~(\ref{eq:diracc}) \& (\ref{eq:dirmloss}).
The accretion/outflow rates are in dimensionless units and are freely scalable.
A vertical green dashed line in each panel marks our standard parameters,
corresponding to $R=1\,\au$ with fiducial field strength and
FUV-penetration depth, and resulting in a wind-driven accretion rate 
$\sim10^{-8}M_{\bigodot}$ yr$^{-1}$. In the left and middle panels, we break the
the dependence of accretion/mass-loss rate on $v_{A0}$ into the dependence on
poloidal field strength $B_{p0}$ and wind-base density $\rho_b$ separately.
The right panel varies the wind temperature ($\propto\csw^2$). 
The middle panel aims to illustrate the role of FUV penetration depth, where deeper
penetration leads to larger wind-base density.

We see that increasing poloidal field strength increases both accretion and mass loss
rates, with the former increasing much faster, corresponding to larger Alfv\'en
radius. Fixing the field strength while enhancing FUV penetration depth (enhanced
$\rho_b$) also leads to both higher accretion and mass-loss rates. In this case, the
system develops a more heavily loaded wind, and winds up toroidal field more (i.e., larger
$B_\phi$), but it is the mass-loss rate that increases faster. Variations in wind
temperature, on the other hand, appear to only weakly affect the wind properties. This is
mainly because the wind base is in pressure balance with the disk interior. For a
steady-state wind, an increase in temperature is compensated by lower density, which does
not alter the relative importance between gas and magnetic pressure (or $v_{A0}/\csw$).

Acknowledging all the uncertainties discussed in \S\ref{ssec:dscp} for our local wind
framework, we expect such an approach to be able to capture the physical essence of
wind-driven evolution of PPDs. We note that a recent effort by \citet{Armitage_etal13} on
PPD evolution adopted wind torques based on scalings from local disk simulations with
turbulence, and neglected disk mass loss. Our wind model have incorporated the physics
at global scales, and the thorough exploration of parameter space allows us to better
account for the controlling effects of disk evolution. In the mean time, the schematic plots
in Figure \ref{fig:maccmloss} highlight the importance of better understanding two issues.

The first is the transport of magnetic flux in PPDs, which determines the
radial profile of poloidal field strength and its evolution. So far, most studies
conform to the framework of inward advection by accretion and outward
transport by turbulent diffusion assuming magnetic field embedded in vacuum
beyond the disk (e.g., \citealp{Lubow_etal94a,Okuzumi_etal14,GuiletOgilvie14}).
The presence of wind plasma with wind-driven accretion may significantly change
the picture \citep{CaoSpruit13}. In addition, non-ideal MHD effects within the disk,
especially the polarity-dependent Hall effect, may also play an important role in
magnetic flux transport, as already hinted by local studies
\citep{Bai14}. On the other hand, regardless of the details of magnetic-flux
transport, the fact that accretion rates of young stellar objects (YSOs) decrease
with age \citep{Fedele_etal10,Sicilia_etal10} implies that the disk must
lose magnetic flux over time, which may permit phenomenological considerations.

The second issue is UV radiative transfer, the associated photochemistry, and their
coupling to wind dynamics, which largely determine the location of the wind base. Despite
increasing sophistication, most photochemical modeling efforts so far have assumed
hydrostatic equilibrium (e.g.,
\citealp{NomuraMillar05,Woitke_etal09,BethellBergin11,Walsh_etal12}), and this is also the
case for FUV photoevaporation studies (e.g., \citealp{Gorti_etal09}).  Density profiles in
the MHD wind-launching region deviate from being purely hydrostatic due to substantial
magnetic pressure support, and asymptote to a $\rho\propto R^{-2}$ profile in the wind,
which differs substantially from Gaussian-like hydrostatic profiles. Consider a situation
with significant wind mass loss (say $\sim10^{-8}M_{\bigodot}$ yr$^{-1}$) with
characteristic velocity of $\sim5$ km s$^{-1}$ (low velocity component) at a
characteristic radius of $\sim1$ AU: the wind column density would reach $\sim0.015$ g
cm$^{-2}$, comparable to the expected penetration depth for far-UV photons. Such shielding
would make wind launching occur at larger vertical height, thus reducing mass
loss, especially towards larger radii. This shielding effect may also directly affect
observational diagnostics of winds, as has been discussed with regard to the survival of
molecules in disk winds \citep{Panoglou_etal12}.

Additional complexities arise from stability considerations. Wind stability has been
discussed in \S\ref{sssec:mechanism}.  Another stability issue concerns the disk itself
when accretion is wind-driven
\citep{Lubow_etal94,CaoSpruit02,Konigl04,Moll12,Lesur_etal13}. Claims for both stability
and instability exist, although these studies have simplified the disk physics
relevant to PPDs, and the situation is far from being clear.  Both wind and disk stability
issues call for realistic 3D simulations.

\subsection[]{Implications for Planet Formation}

Disk winds may affect many aspects of planet formation by
influencing the global evolution of PPDs through angular-momentum transport
and mass loss. The fact that we expect PPDs to lose mass quickly has a
more direct consequence on planet formation: it will promote planetesimal
formation.

Planetesimal formation remains the least understood process in the theory of planet
formation (see review by \citealp{ChiangYoudin10}). Among possible mechanisms, streaming
instability \citep{GoodmanPindor00,YoudinGoodman05} appears to be the most
promising. However, to form planetesimals via the streaming instability, the dust-to-gas
ratio must be enhanced by a factor of several above solar abundance, $\sim0.01$ 
\citep{Johansen_etal07,Johansen_etal09,BaiStone10b,BaiStone10c}.  This requirement poses a
strong challenge for planetesimal formation in the terrestrial-planet-forming region
within the snow line \citep{DrazkowskaDullemond14}.

In the presence of a disk wind, gas is removed from several scale heights
above disk midplane. With the disk being largely laminar or weakly turbulent,
most dust grains are expected to settle toward the disk midplane. Therefore,
we expect the wind to primarily remove gas with depleted dust content, effectively
enhancing the dust-to-gas mass ratio. \citet{Gorti_etal15} explored this idea
within the framework of photoevaporation and found that substantial enhancement
dust-to-gas ratios can be achieved. We expect similar conclusions to
hold in the MHD-wind framework, and this effect may be even more prominent
given the further enhanced mass-loss rate. Therefore, planetesimal formation
via the streaming instability may become feasible in the inner region of PPDs.

Using CO as a gas tracer, \citet{WilliamsBest14} reported evidence of
enhanced dust-to-gas ratios for a group of PPDs. Their results mainly apply
to the outer parts of disks, which contain the bulk of the mass reservoir. Although
the reliability of their gas-mass estimates remains questionable, the estimates
suggest that wind mass loss may also be significant in the outer disk.

\subsection[]{Connection to Black Hole Accretion Disks}

While we have focused on magnetized disk winds from PPDs, the results
presented here may also shed light on accretion disks surrounding compact
objects, especially black holes (BHs). While BH accretion disks are fully MRI
turbulent, local MHD simulations show that wind launching in the presence of
net vertical magnetic flux typically occurs at several scale heights above the
disk midplane \citep{SuzukiInutsuka09,Fromang_etal13}. Global simulations
of hot accretion flows have also identified magnetic forces for driving disk
outflows \citep{Yuan_etal12,Yuan_etal15}. Moreover, magnetic
fields become more ordered as net vertical field becomes stronger
\citep{BaiStone13a}, permitting a quasi-steady-state model construction.

In the context of active galactic nuclei (AGN) disks hosting supermassive BHs,
powerful winds are commonly observed in broad absorption lines, with inferred
mass loss rates comparable to or even higher than accretion rates
(see \citealp{Crenshaw_etal03} for a review). While radiation force likely plays
a dominant role in many cases for wind driving, not all AGN winds can be
explained by radiation driving alone (see \citealp{Proga07} for a review),
and magnetic driving also appears to be essential
\citep{KoniglKartje94,deKoolBegelman95,Everett05,Kraemer_etal05}. In the
context of X-ray binaries hosting stellar mass black holes, disk wind signatures
are commonly observed in the disk-dominated soft states
(e.g., \citealp{NeilsenLee09,King_etal12,Miller_etal12,Ponti_etal12}), again with
mass loss rate comparable to accretion rate. While very high degree of ionization
requires strong X-ray illumination,
photoionization modeling shows that neither radiation nor thermal pressure can
account for the wind kinematics, hence magnetic driving is inevitable (e.g.,
\citealp{Miller_etal06,NeilsonHoman12}). In both contexts, provided the estimates
for the size of the wind launching region, wind temperature, together with wind
velocity and mass flux based on photoionization modeling, it is possible to place
useful constraints on disk magnetization from our wind framework. This parameter
is of great theoretical interests, and likely plays a decisive role in overall disk evolution.

\section[]{Summary and Conclusions}\label{sec:sum}

In this paper, we have constructed a simple 1D model for the kinematics of
magnetized winds launched from PPDs. It is motivated by local simulations of
PPD gas dynamics which included detailed microphysics, as well as theoretical
and observational studies of PPD photoevaporation. We aim to unite the two
scenarios of PPD mass loss into a joint theoretical framework, which we coin
magneto-photoevaporation or magneto-thermal winds, since both magnetic
and thermal effect play a major role in wind launching and kinematics.
Key ingredients in this model are the following:
\begin{itemize}
\item Wind launching takes place in the warm disk surface that is well
ionized by (far-) UV radiation, while the cold and poorly ionized disk interior
does not participate (see Figure \ref{fig:illus}).
\item The geometry and strength of poloidal magnetic field lines are prescribed,
as is thermodynamics (sound speed); these serve as main parameters.
\item The model is built upon conservation laws in axisymmetric steady state
and ideal MHD, and is a direct generalization of the \citet{WeberDavis67} wind.
\end{itemize}

We choose our fiducial wind model to mimic wind launched from $\sim1\,\au$
with a field strength that would drive an accretion rate
$\sim10^{-8}M_{\bigodot}$ yr$^{-1}$.  But the model scalable, and we have made a
thorough exploration of the parameter space surrounding the fiducial model. Most of our investigation has
focused on the wind mass-loss rate and wind-driven accretion rate, whose
ratio is directly connected to the Alfv\'en radius [see eq.~\eqref{eq:mdot2}].
The main findings are best summarized in Figure \ref{fig:maccmloss},
including these:
\begin{itemize}
\item Increasing the poloidal field strength rapidly increases the accretion
rate, and less rapidly the mass-loss rate, making the wind more
lightly loaded.
\item Increasing the FUV penetration depth rapidly increases the
mass-loss rate, and less rapidly the mass accretion rate, making the
wind more heavily loaded.
\item Increasing only the wind temperature modestly affect boths the accretion
rate and mass loss rate.
\end{itemize}
In essence, the wind kinematics is largely determined by the ratio of poloidal Alfv\'en
velocity $v_{A0}$ and sound speed at the wind base $\csw$ (see Figure
\ref{fig:thermal}). Strongly magnetized winds ($v_{A0}>\csw$) have larger Alfv\'en radii
and are lightly loaded. We also find that the classical centrifugal driving applies only
when $v_{A0}\gg \csw$; otherwise the wind driving mechanism is largely due to gradients in
the toroidal magnetic pressure as is most likely the case for PPD winds. This fact raises
concerns about the stability of PPD winds that demand further investigation.

We find that the general wind properties are not very sensitive to the geometric shape of
the field lines, nor to the modest level of sub-Keplerian or super-Keplerian disk
rotation. However, they do depend relatively strongly on the poloidal field strength and
how rapidly $B_p$ achieves a $R^{-2}$ rather than $R^{-1}$ profile.  This in turn may
depend on how poloidal magnetic flux is distributed in the disk. We expect radially
concentrated (extended) flux distributions to yield relatively smaller (larger) $R_A$ and
greater (less) local mass loading.  Yet the difference in global mass-loss rate may not
be so dramatic because the more heavily loaded winds may be confined to smaller radial
ranges by their more concentrated flux distributions.

In addition, we find that most kinematic scaling laws in the cold
Weber \& Davis wind model can be extended to PPD disk winds, with certain
renormalizations, as long as the wind is relatively cold ($\csw/v_K\lesssim0.1$).

Overall, our results suggest that due to the dominant role played by
magnetic fields, PPDs are likely lose mass more quickly than previously
thought (from pure photoevaporation), and reach a considerable fraction
of the wind-driven accretion rate. Bearing in mind the uncertainties
and caveats, the connection between wind-driven accretion and wind
mass loss and their dependence on disk parameters that we have studied
in this work may facilitate future development toward a realistic theoretical
framework for global PPD structure and evolution.

\acknowledgments

We thank Uma Gorti, Geoffroy Lesur and Eve Ostriker for useful discussions.
XNB acknowledges support from Institute for Theory and Computation (ITC)
at Harvard-Smithsonian Center for Astrophysics.  JG acknowledges support
from the NASA Origins of Solar Systems program via grant NNX10AH37G.
FY acknowledges support by the NSF of China (grants 11133005 and 11573051),
and the Strategic Priority Research Program ÒThe Emergence of Cosmological
StructuresÓ of CAS (grant XDB09000000).

\appendix

\section[]{A. Estimate of Fiducial Wind Parameters}\label{app:bstr}

PPDs show signatures of active accretion onto the protostar, with typical
accretion rate on the order of $10^{-8}M_{\bigodot}$ yr$^{-1}$
\citep{Hartmann_etal98}, which requires efficient transport of disk angular
momentum.  As discussed in \citet{Wardle07} and \citet{BaiGoodman09}, this fact
alone places an important constraint on the magnetic field strength in PPDs. For
wind-driven accretion, accretion rate is directly proportional to the magnetic torque
exerted at the disk surface $\dot{M}\approx2|B_zB_\phi|R/\Omega$, and
correspondingly, the expected field strength at disk surface is
\begin{equation}
B_{p0}\approx0.065G\cdot\dot{M}_{-8}^{1/2}
\bigg(\frac{B_p}{B_\phi\sin\theta}\bigg)^{1/2}R_{\au}^{-5/4},\label{eq:Bp0}
\end{equation}
for disks orbiting a solar mass star, where $\dot{M}_{-8}$ is disk accretion rate
normalized to $10^{-8\pm1}M_{\bigodot}$ yr$^{-1}$, and $R_{\au}$ is the
cylindrical radius normalized to 1 AU. We note that transport of angular
momentum vertically by disk wind is more efficient than radially by Maxwell
stress by a factor of $\sim R/H$. Therefore, for a given accretion rate, the required
field strength for wind-driven accretion is weaker. Recently, \citet{Fu_etal14} reported
the magnetic field strength of the early solar nebular deciphered from the Semarkona
meteorite. Within uncertainties, the inferred (midplane) field strength is relatively
weak and is more compatible with the wind-driven accretion scenario.

To non-dimensionize this field strength at the wind base for our model, we apply
the standard minimum-mass solar nebular (MMSN) disk model
\citep{Weidenschilling77b,Hayashi81} with
\begin{equation}
\Sigma_{\rm MMSN}=1700R_{\au}^{-3/2}\ {\rm g\ cm}^{-2}\ ,\qquad
T=280R_{\au}^{-1/2}\ {\rm K}\ .
\end{equation}
Here, disk temperature is assumed to be vertically isothermal, and it refers to the ``cold"
region marked in Figure \ref{fig:illus}, e.g., disk interior. The normalized disk scale height is
given by $H/R=c_s/v_K\approx0.034R_{\au}^{1/4}$, where $c_s$ is the isothermal
sound speed in the disk interior.

We first discuss the location of the wind base $z_0$, corresponding to the FUV ionization
front. Assuming a FUV penetration of depth
of $\gtrsim0.01$ g cm$^{-2}$ \citep{PerezBeckerChiang11b} and hydrostatic equilibrium,
the wind base location is approximately at $z=4.4H$ for $R_0=1$ AU. This
is further complicated by two factors. First, the gas in the wind zone is expected to be denser
than from hydrostatic equilibrium, and hence the FUV ionization front should be located
higher. Local isothermal wind simulations of \citet{BaiStone13b} find $z_0\sim4.6H$. Second,
the gas in the wind zone is heated to become less dense, which counter-balances the
previous effect to bring the FUV ionization front lower. Based on these considerations,
we place $z_0$ at around $\sim4.5H$ for wind launched from $R_0=1$ AU.
For wind launched from other radii, assuming constant FUV penetration depth, we expect
$z_0/H$ be smaller toward larger radii (if shielding of FUV by the wind flow launched from
smaller radii is negligible).

Hydrostatic equilibrium applies in the disk interior within $z=\pm z_0$. The gas density on
the disk side of the wind base is
\begin{equation}
\rho_{b0}\approx1.4\times10^{-9}\bigg(\frac{\Sigma}{\Sigma_{\rm MMSN}}\bigg)R_{\au}^{-11/4}
\exp{\bigg[-\frac{1}{2}\bigg(\frac{z_0}{H}\bigg)^2\bigg]}\ {\rm g\ cm}^{-3}\ .
\end{equation}
Using Equation (\ref{eq:Bp0}), we obtain the plasma $\beta$ for the poloidal field at wind base
\begin{equation}
\beta_{p0}=\frac{\rho_{b0}c_s^2}{B_{p0}^2/8\pi}=2\frac{c_s^2}{v_{Ap}^2}\approx8.2\times10^4
\dot{M}_{-8}^{-1}\exp{\bigg[-\frac{1}{2}\bigg(\frac{z_0}{H}\bigg)^2\bigg]}
\bigg(\frac{\Sigma}{\Sigma_{\rm MMSN}}\bigg)
\bigg(\frac{B_p}{B_\phi\sin\theta}\bigg)^{-1}R_{\au}^{-3/4}\ ,\label{eq:va0par}
\end{equation}
where $v_{Ap}$ is the poloidal Alfv\'en velocity on the disk side of the wind base.
Despite the jump of gas temperature across the wind base into the wind zone,
$\beta_{p0}$ is unaltered across the wind base, because both $B_p$ and
gas pressure must vary smoothly. Therefore, the ratio $v_{Ap}/c_s=v_{A0}/\csw$
is unaltered across the wind base ($v_{A0}, \csw$ are the counterparts of 
$v_{Ap}, c_s$ in the wind zone).

Taking $z_0=4.5H$, we find $\beta_{p0}\approx3.3$ and hence
$v_{A0}\approx0.8\csw$ assuming all other dimensionless ratios are of order unity.
This number should be taken as an order of magnitude estimate, given the
super-exponential dependence on $z_0$ as well as uncertainties in other parameters.
Therefore, we choose $v_{A0}=\csw$ in our fiducial calculations, while also
explore a number of variations with $v_{A0}/\csw\sim0.3-10$.

In our calculations, velocities are normalized to $v_K$. At the fiducial radius of $R=1$ AU,
gas temperature rapidly increases from $\lesssim300$K in the disk interior to a few
thousand K into the FUV layer (e.g., \citealp{Walsh_etal10}). Fiducially, we consider a factor
of $\sim10$ increase in gas temperature across the wind base/FUV ionization front. This
corresponds to a factor $\sim3$ increase in sound speed, giving
$\csw\approx3c_s\approx0.1v_K$ at 1 AU. We also explore a much wider range of
$\csw$ in our calculations.

\bibliographystyle{apj}

\label{lastpage}
\end{document}